\documentstyle[epsfig]{new}
\begin{document}

\title{%
Three flavor neutrino oscillation analysis of
the Superkamiokande atmospheric neutrino data}

\author{Osamu Yasuda \\
{\it Department of Physics, Tokyo Metropolitan University,\\
1-1 Minami-Osawa Hachioji, Tokyo 192-0397, Japan, \\
yasuda@phys.metro-u.ac.jp}}

\maketitle

\section*{Abstract}

Superkamiokande atmospheric neutrino data for 414 days are analyzed in
the framework of three flavor oscillations with mass hierarchy.  It is
shown that the best fit point is very close to the pure maximal
$\nu_\mu\leftrightarrow\nu_\tau$ case and $\Delta m^2 \simeq
7\times10^{-3}$ eV$^2$.  The allowed region at 90 \%CL is given and
the implications to the long baseline experiments are briefly
discussed.  It is also shown that the threefold maximal
mixing model fits to the data of atmospheric neutrinos and
the reactor experiments for $\Delta m^2 \sim 1\times10^{-3}$ eV$^2$.

\section{Introduction}

Recent atmospheric neutrino data of the Superkamiokande experiment
$^{11,12,17}$ provide very strong evidence for neutrino oscillations.
The original analysis by the Superkamiokande group is based on the two
flavor framework.  Despite the constraints from the reactor
experiments$^{27,1,2}$, there can be small mixing between $\nu_\mu$
and $\nu_e$ in the atmospheric neutrinos.  To estimate the oscillation
probability of $\nu_e\leftrightarrow\nu_x$ or
$\bar\nu_e\leftrightarrow\bar\nu_x$ in the long baseline
experiments$^{23,21,16,25,26}$, therefore, it is important to know the
allowed region of the mixing parameters at a certain confidence level.
In this talk I present a three flavor analysis of the Superkamiokande
atmospheric neutrino data for 414 days$^{11}$.  Some of the contents
of this presentation overlaps with Lisi's talk$^{18,7}$.

\section{Three flavor analysis of the Superkamiokande 
atmospheric neutrino data}

To evaluate the number of events, we integrate numerically the
Schr\"odinger equation
\begin{eqnarray}
i {d \over dx} \left( \begin{array}{c} \nu_e (x) \\ \nu_{\mu}(x) \\ 
\nu_{\tau}(x)
\end{array} \right) = 
\left[ U {\rm diag} \left(0,\Delta E_{21},\Delta E_{31} \right) U^{-1}
+{\rm diag} \left(A,0,0 \right) \right]
\left( \begin{array}{c} \nu_e (x) \\
\nu_{\mu}(x) \\ \nu_{\tau}(x)
\end{array} \right),
\label{eqn:sch}
\end{eqnarray}
where
\begin{eqnarray}
U\equiv\left(
\begin{array}{ccc}
U_{e1} & U_{e2} &  U_{e3}\\
U_{\mu 1} & U_{\mu 2} & U_{\mu 3} \\
U_{\tau 1} & U_{\tau 2} & U_{\tau 3}
\end{array}\right)\nonumber
\end{eqnarray}
is the MNS mixing matrix$^{19}$, $E$ is the neutrino energy,
$A\equiv\sqrt{2} G_F N_e(x)$ stands for the matter effect$^{20,28}$ in the
Earth, $\Delta E_{ij}\equiv\Delta m^2_{ij}/2E$, $\Delta
m^2_{ij}\equiv m^2_i-m^2_j$.  We assume without loss of generality
that $\Delta m^2_{21}$ has the smallest absolute value of the mass
squared difference.  To account for the solar neutrino deficit$^{3}$
we assume that $\Delta m^2_{21}$ is of order $10^{-5}$eV$^2$ or
$10^{-10}-10^{-9}$eV$^2$. On the other hand, the zenith angle
dependence in the atmospheric neutrino anomaly requires that the
larger mass squared difference be of order $10^{-3}-10^{-2}$eV$^2$.
So the pattern of the mass squared differences has hierarchy and can
be classified into two cases which are depicted in Fig. 1.  \vglue -
1.5cm\hglue 1.5cm \epsfig{file=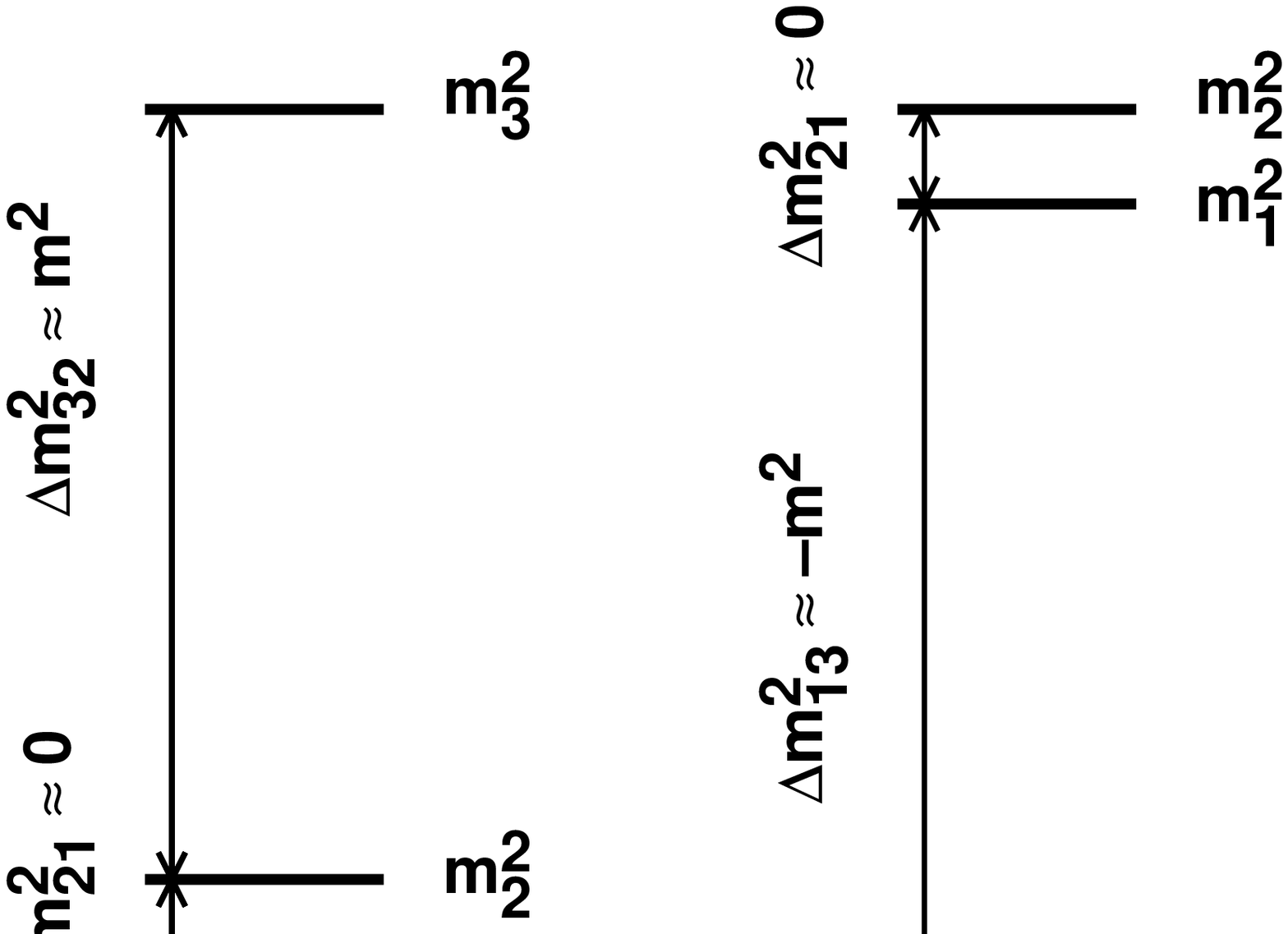,width=8cm}

\vglue 3.0cm\hglue 0cm\parbox{13cm}
{Fig. 1~~The hierarchical neutrino mass squared differences.
The scenarios (a) and (b) are related by exchanging
$m^2\leftrightarrow-m^2$.  They are equivalent in vacuum but
physically inequivalent in matter.}

The eigenvalues of the Schr\"odinger equation (\ref{eqn:sch})
are given by$^4$

\begin{eqnarray}
\left\{ \begin{array}{lll}
{\alpha /3}&+&(2/3)\sqrt{\alpha^2-3\beta}\cos\left(\chi / 3\right)\\
{\alpha / 3}&-&(2/3)\sqrt{\alpha^2-3\beta}\cos\left((\chi-\pi)/3\right)\\
{\alpha / 3}&-&(2/3)\sqrt{\alpha^2-3\beta}\cos\left((\chi+\pi)/3\right),
             \end{array} \right.
\label{eqn:eigen}
\end{eqnarray}
where
\begin{eqnarray}
\chi&\equiv&\cos^{-1}\left[{2\alpha^3-9\alpha\beta+27\gamma \over
2(\alpha^2-3\beta)^{3/2}}\right]\nonumber\\
\alpha&\equiv&\Delta E_{31}+\Delta E_{21}+A\nonumber\\
\beta&\equiv&\Delta E_{31}\Delta E_{21}+A\left[\Delta E_{31}(1-|U_{e3}|^2)
+\Delta E_{21}(1-|U_{e2}|^2)\right].\nonumber\\
\gamma&\equiv&\Delta E_{31}\Delta E_{21}A\,|U_{e1}|^2\nonumber
\end{eqnarray}
Under our assumption that $\Delta m^2\lsim 10^{-5}$eV$^2$, in the
contained atmospheric neutrino events (0.1 GeV $\le$ $E$ $\le$ 100
GeV) we have $|\Delta E_{21}L|\ll 1$ for downward going events ($L$ is
the neutrino path length and $L\lsim$ 500 km), and $|\Delta E_{31}|$,
$|\Delta E_{32}|$, $A$ $\gg$ $|\Delta E_{21}|$ for upward going
events.  Thus the eigenvalues in (\ref{eqn:eigen}) with $\Delta
m^2_{21}\sim{\cal O}$($10^{-5}$eV$^2$) or ${\cal O}$($10^{-10}$eV$^2$)
are almost the same as those with $\Delta m^2_{21}\equiv 0$ to good
precision, i.e., {\it the matter effects are much more important than
the smaller mass squared difference $\Delta m_{21}^2\lsim 10^{-5}$eV$^2$
in the atmospheric neutrino
problem.} Hence we can safely ignore $\Delta m^2_{21}$ in the
following discussions.  In the limit $\Delta m^2_{21}=0$ the mass
matrix on the right hand side of (\ref{eqn:sch}) becomes
\begin{eqnarray}
D
\left[ U(\theta_{12}=\delta=0) {\rm diag} \left(0,0,\Delta E_{31}\right)
U^{-1}(\theta_{12}=\delta=0)
+{\rm diag} \left(A,0,0 \right) \right]
D^{-1},
\label{eqn:sch2}
\end{eqnarray}
where $D\equiv {\rm diag}\left(e^{-i\delta},1,1 \right)$, we have used
the standard parametrization$^{24}$ of the MNS mixing matrix$^{19}$
\begin{eqnarray}
U\equiv e^{i\theta_{23}\lambda_7}
{\rm diag}\left(e^{-i\delta/2},1,e^{i\delta/2} \right)
e^{i\theta_{13}\lambda_5}
{\rm diag}\left(e^{i\delta/2},1,e^{-i\delta/2} \right)
e^{i\theta_{12}\lambda_2}\nonumber
\end{eqnarray}
and
\begin{eqnarray}
\lambda_2=\left(
\begin{array}{ccc}
0&-i& 0\\
i&0&0\\
0&0&1
\end{array}\right),
~\lambda_5=\left(
\begin{array}{ccc}
0&0& -i\\
0&1&0\\
i&0&0
\end{array}\right),
~\lambda_7=\left(
\begin{array}{ccc}
1&0& 0\\
0&0&-i\\
0&i&0
\end{array}\right),\nonumber
\end{eqnarray}
are the Gell-Mann matrices.  As we see in (\ref{eqn:sch2}),
$\theta_{12}$ drops out and so does $\delta$, as $\delta$
appears only in the overall phase of the oscillation amplitude
$A(\nu_\alpha\rightarrow\nu_\beta)$.  Putting $m^2\equiv\Delta m^2_{31}$,
therefore, we are left with the three parameters 
($m^2, \theta_{13}, \theta_{23}$).

The way to obtain the numbers of events is exactly the same as in Foot
et al.$^8$, and we refer to Foot et al.$^8$ for details.  In
Foot et al.$^8$ two quantities have been introduced to perform a $\chi^2$
analysis.  One is the double ratio$^{15,10}$
\begin{eqnarray} 
R \equiv \frac{(N_{\mu}/N_e)|_{\rm osc}}{(N_{\mu}/N_e)|_{\rm no-osc}}
\nonumber
\end{eqnarray}
where the quantities $N_{e,\mu}$ are the numbers of $e$-like and
$\mu$-like events. The numerator denotes numbers with oscillation
probability obtained by (\ref{eqn:sch}), while the denominator the
numbers expected with oscillations switched off.  The other one is the
quantity on up-down flux asymmetries for $\alpha$-like
($\alpha$=e,$\mu$) events and is
defined by
\begin{eqnarray}
Y_{\alpha} \equiv {(N_{\alpha}^{-0.2}/N_{\alpha}^{+0.2})|_{\rm osc}
\over (N_{\alpha}^{-0.2}/N_{\alpha}^{+0.2})|_{\rm no-osc}},\nonumber
\end{eqnarray}
where $N_{\alpha}^{-\eta}$ denotes the number of $\alpha$-like events 
produced in
the detector with zenith angle $\cos \Theta < -\eta$, while
$N_{\alpha}^{+\eta}$ denotes the analogous quantity for $\cos \Theta >
\eta$, where $\eta$ is defined to be positive.  Superkamiokande divides the
$(-1,+1)$ interval in $\cos\Theta$ into five equal bins, so we choose
$\eta = 0.2$ in order to use all the data in the other four bins.
Thus $\chi^2$ for atmospheric neutrinos is defined by
\begin{eqnarray}
\chi^2_{\rm atm} = \sum_E \left[\left({R^{SK} - R^{th} \over
\delta R^{SK}}\right)^2
+ \left({Y^{SK}_{\mu} - Y^{th}_{\mu} \over \delta Y^{SK}_{\mu}}\right)^2
+ \left({Y^{SK}_{e} - Y^{th}_{e} \over \delta Y^{SK}_{e}}\right)^2
\right],\nonumber
\end{eqnarray}
where the sum is over the sub-GeV and multi-GeV cases, the measured
Superkamiokande values and errors are denoted by the superscript
``SK'' and the theoretical predictions for the quantities are labeled
by ``th''.  In Foot et al.$^8$ a $\chi^2$ analysis has been performed using
the quantities R and Y's, or using Y's only.  Throughout this paper we
use the quantities R and Y's to get narrower allowed regions for the
parameters.  We have to incorporate also the results of the reactor
experiments.  We define the following $\chi^2$:
\begin{eqnarray}
\chi^2_{\rm reactor}=
\sum_{j=1,12}^{\rm CHOOZ} \left( {x_j - y_j \over \delta x_j}
\right)^2+
\sum_{j=1,60}^{\rm Bugey} \left( {x_j - y_j \over \delta x_j}
\right)^2+
\sum_{j=1,8}^{\rm Krasnoyarsk} \left( {x_j - y_j \over
\delta x_j} \right)^2,\nonumber
\end{eqnarray}
where $x_i$ are experimental values and $y_i$ are the corresponding
theoretical predictions, and the sum is over 12, 60, 8 energy bins of
data of CHOOZ$^2$, Bugey$^1$ and Krasnoyarsk$^{27}$, respectively.
There are 6 atmospheric and 80 reactor pieces of data in
$\chi^2\equiv\chi^2_{\rm atm}+\chi^2_{\rm reactor}$ and 3 adjustable
parameters, $m^2$, $\theta_{13}$ and $\theta_{23}$, leaving 83 degrees
of freedom.

The best fit is obtained for ($m^2, \tan^2\theta_{13},
\tan^2\theta_{23}, \chi^2) = (7\times10^{-3}$ eV$^2$,
$1.1\times10^{-2}$, 0.93, 71.8) for $m^2>0$ and ($-7\times10^{-3}$
eV$^2$, $1.5\times10^{-2}$, 1.1, 71.4) for $m^2<0$, respectively.
$\chi^2_{\rm min}=\left(\chi^2_{\rm atm}\right)_{\rm min}
+\left(\chi^2_{\rm reactor}\right)_{\rm min}=4.2+67.2=71.4$ indicates
that a fit to data is good for 83 degrees of freedom at the best fit
point.  The allowed region for $m^2$ with $\theta_{13}$, $\theta_{23}$
unconstrained is given in Fig. 2, where $\Delta\chi^2\equiv
\chi^2_{\rm atm}+\chi^2_{\rm reactor}
-\left(\chi^2_{\rm atm}+\chi^2_{\rm reactor}\right)_{\rm min}
<3.5,~6.3,~11.5$ corresponds to $1\sigma$, 90 \% CL and 99 \% CL,
respectively.  The allowed region for $|m^2|$ at 99\% CL is
$4\times10^{-4}$ eV$^2$ $\lsim |m^2| \lsim 1.7\times10^{-2}$ eV$^2$.
\vglue -0.5cm\hglue -1.0cm
\epsfig{file=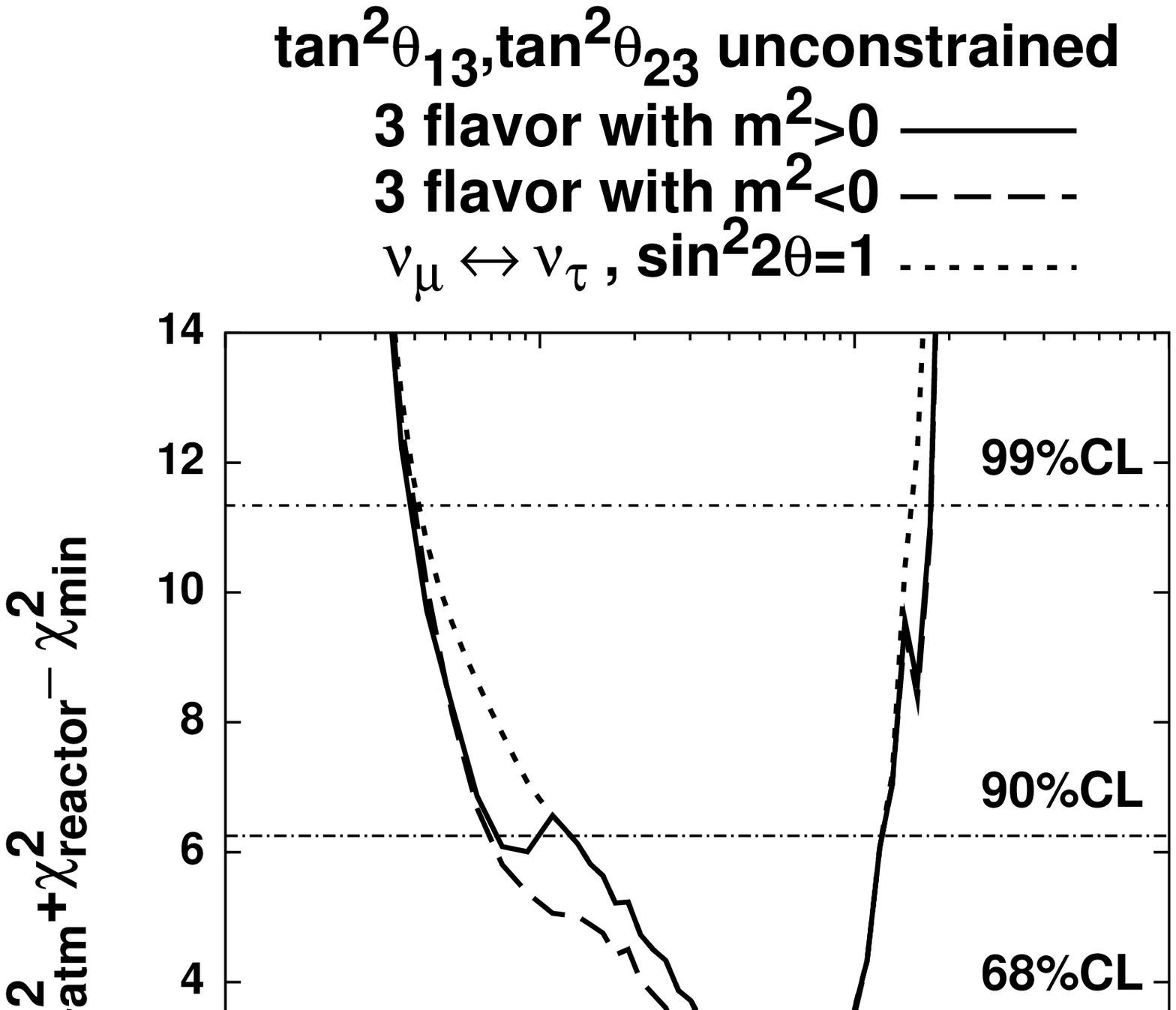,width=8cm}
\vglue -3.5cm\hglue 8cm
\parbox{5cm}{Fig. 2\\
\fussy{\small Value of $\Delta\chi^2\equiv
\chi^2_{\rm atm}+\chi^2_{\rm reactor}
-\left(\chi^2_{\rm atm}+\chi^2_{\rm reactor}\right)_{\rm min}
=\chi^2_{\rm atm}+\chi^2_{\rm reactor}-71.4$.  The
solid, dash-dotted, dotted lines represent the scenarios (a), (b) and
the two flavor case with maximal $\nu_\mu\leftrightarrow\nu_\tau$
mixing, respectively.}}
\vglue 4.5cm
Using the same parametrization as that in Fogli et al.$^5$, the results for
the allowed region of the mixing angles ($\theta_{13}$, $\theta_{23}$)
are given for various values of $m^2$ in Figs. 3 and 4.  The results
for $m^2>0$ (Fig. 3) and $m^2<0$ (Fig. 4) are almost the same.  Notice that
$\theta_{13}$ can be large
in the region of $m^2$ where the CHOOZ data$^2$ give no constraint
on $\theta_{13}$.
\newpage
\vglue 2.5cm
\hglue -1cm 
\epsfig{file=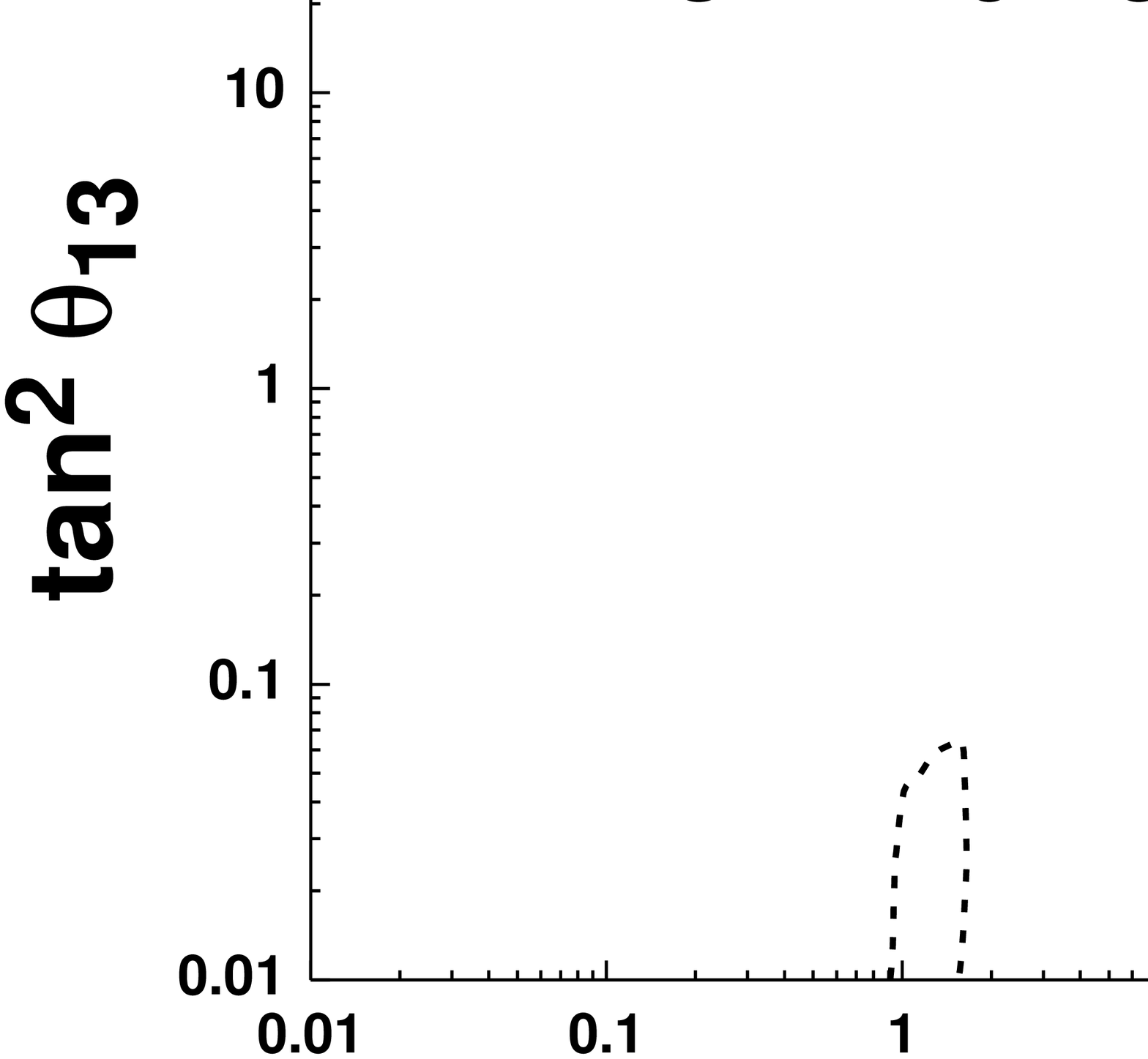,width=3.5cm}
\vglue -3.1cm \hglue 4.3cm \epsfig{file=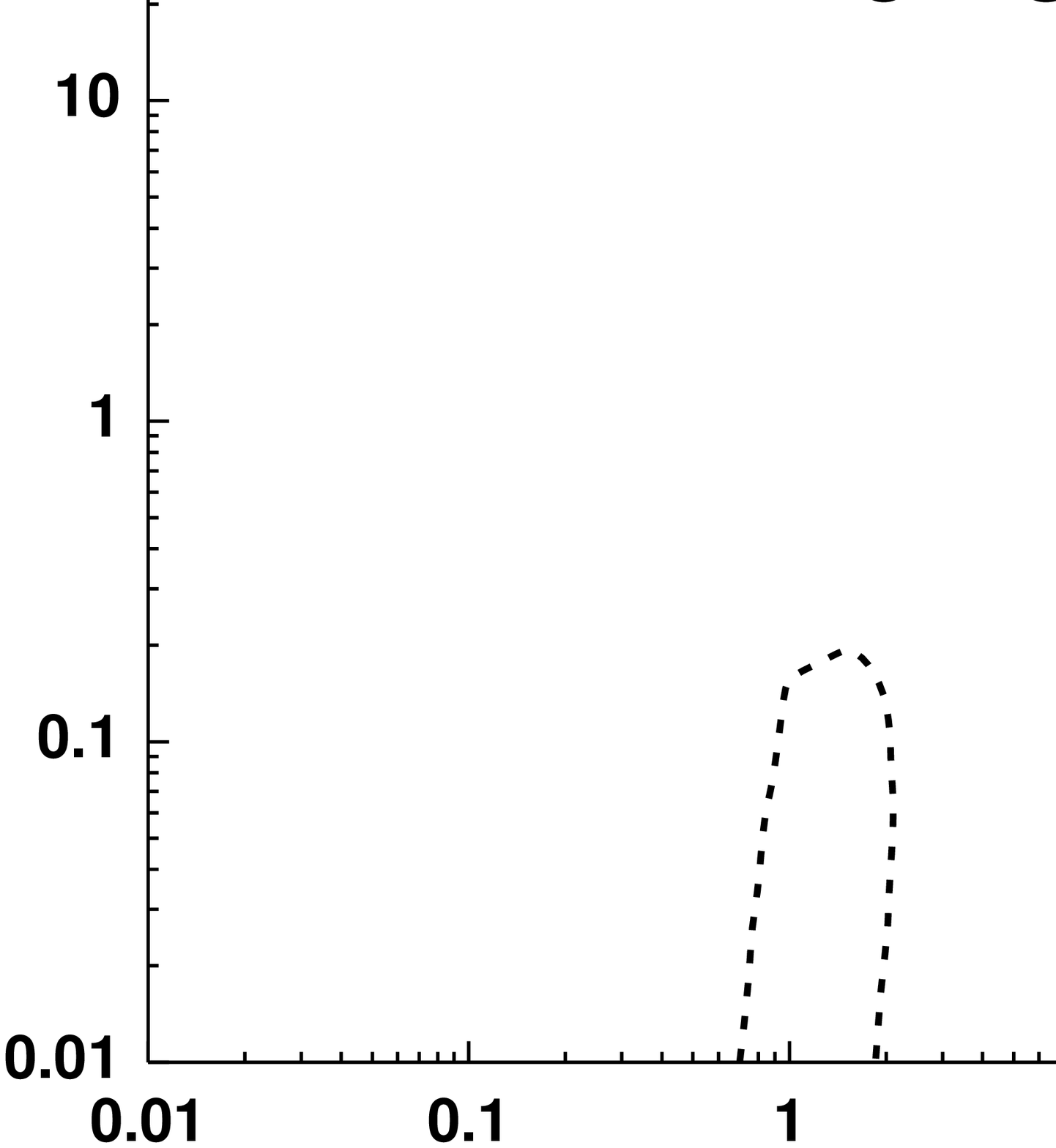,width=3.5cm}
\vglue -3.1cm \hglue 9.6cm \epsfig{file=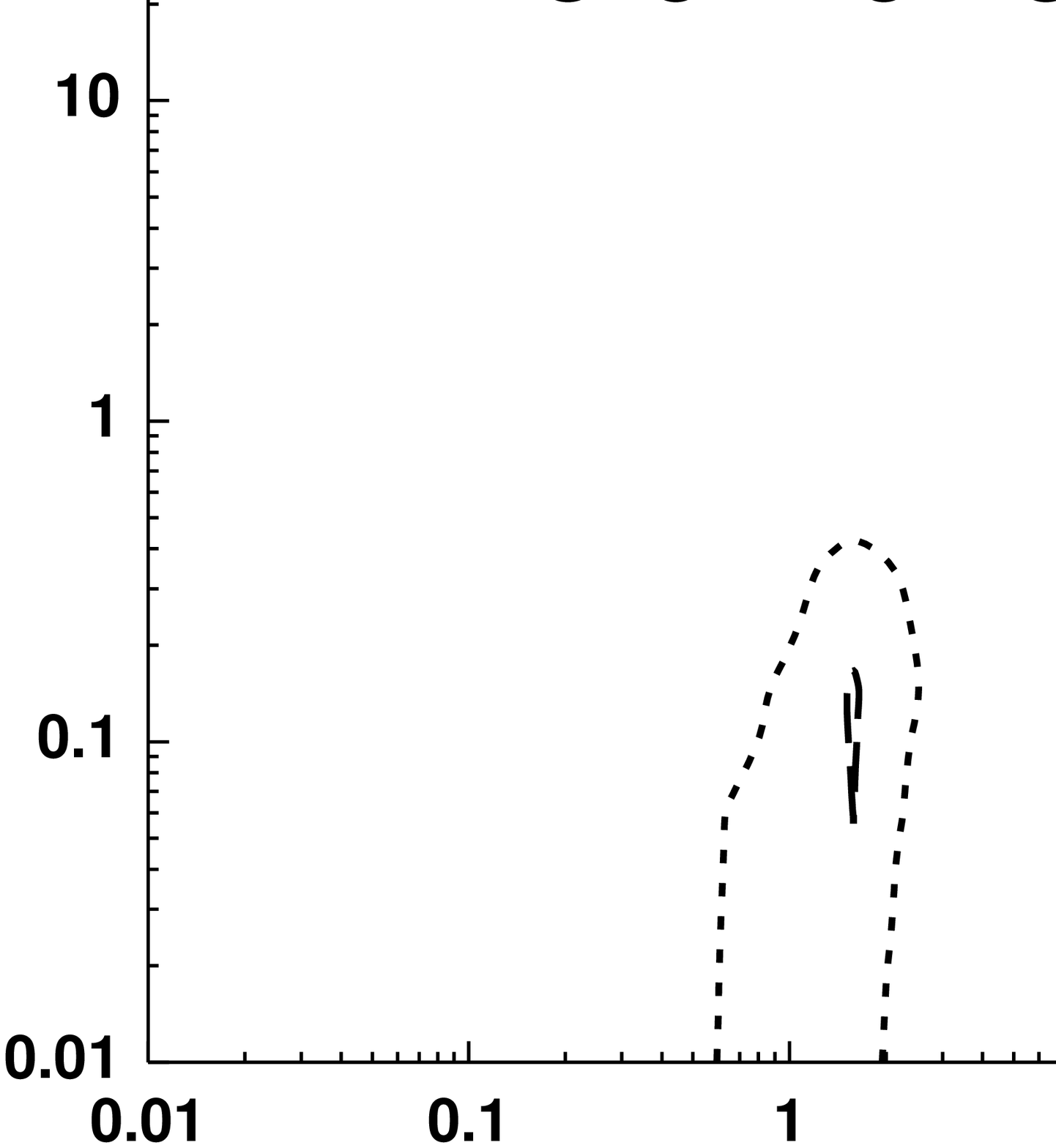,width=3.5cm}

\vglue 1.8cm
\hglue -1cm 
\epsfig{file=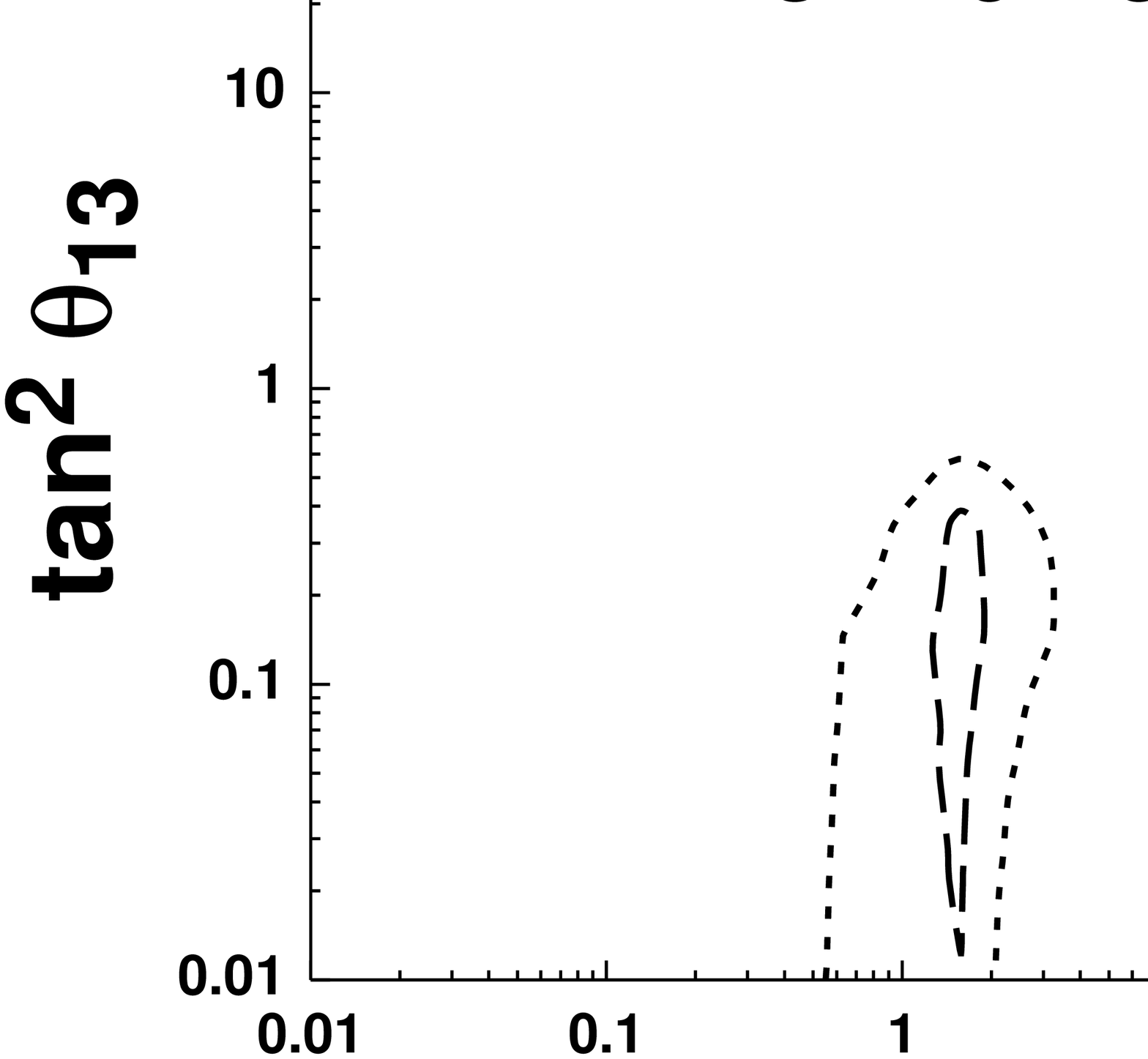,width=3.5cm}
\vglue -3.1cm \hglue 4.3cm \epsfig{file=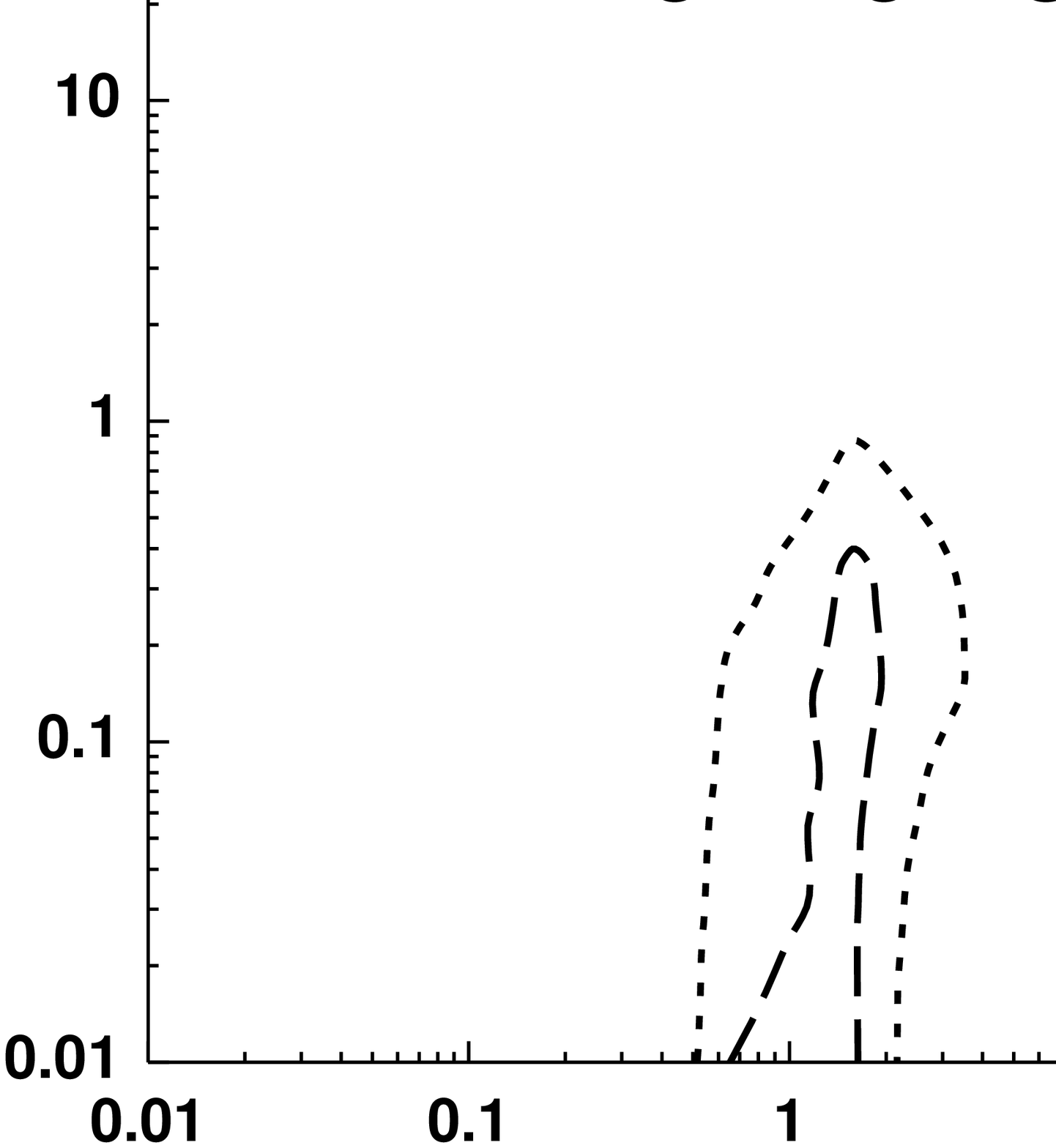,width=3.5cm}
\vglue -3.1cm \hglue 9.6cm \epsfig{file=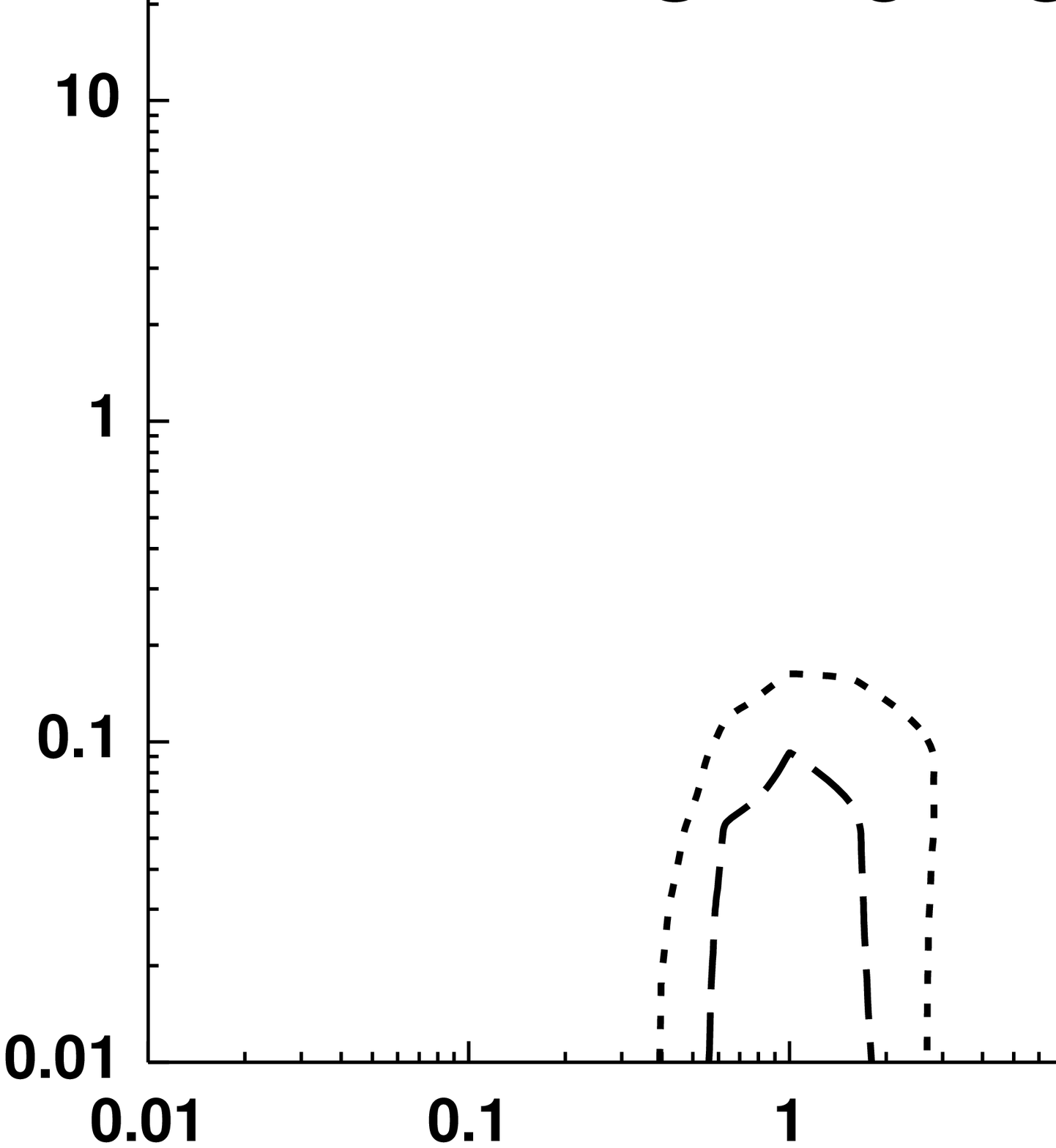,width=3.5cm}

\vglue 1.8cm
\hglue -1cm 
\epsfig{file=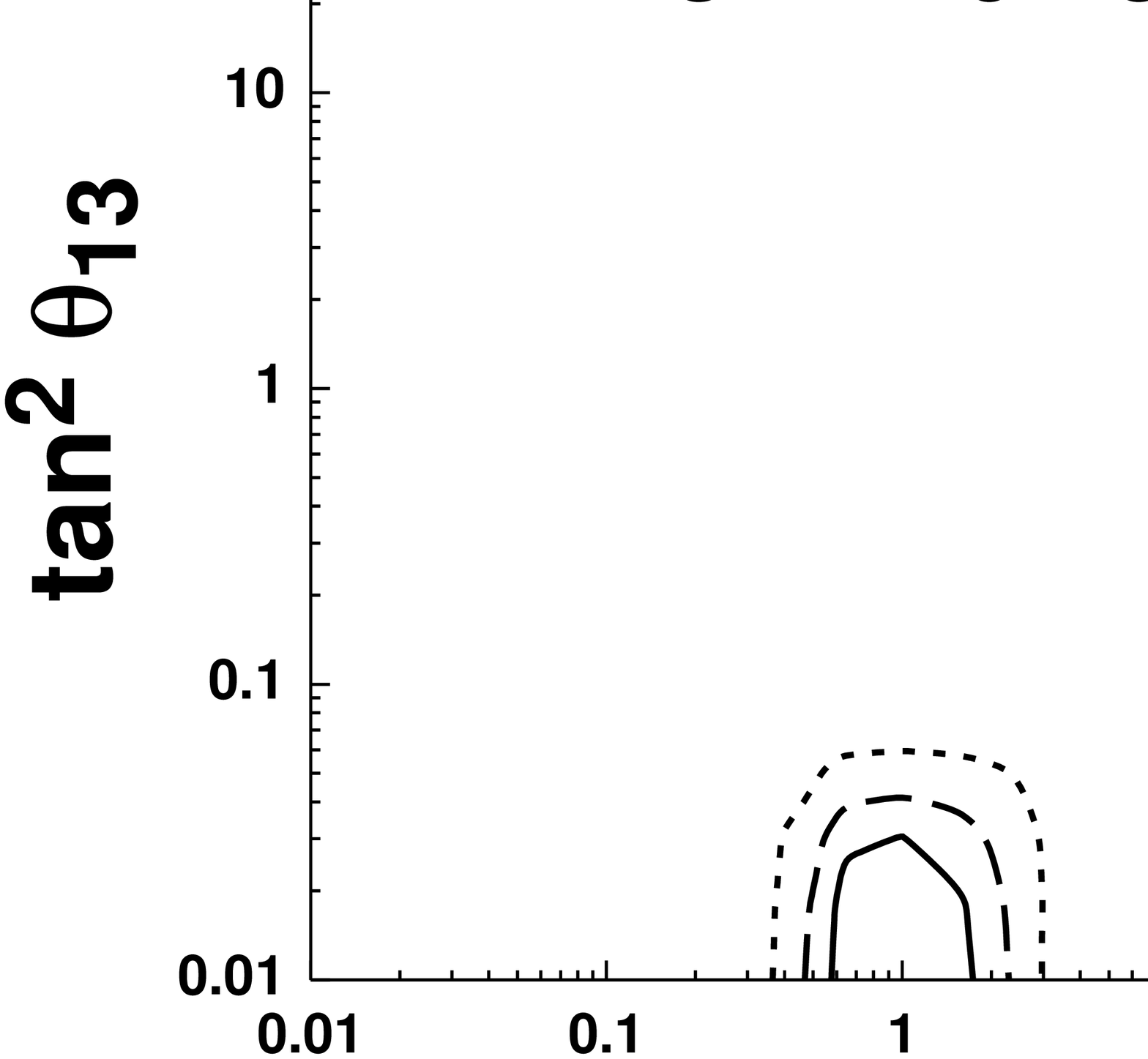,width=3.5cm}
\vglue -3.1cm \hglue 4.3cm \epsfig{file=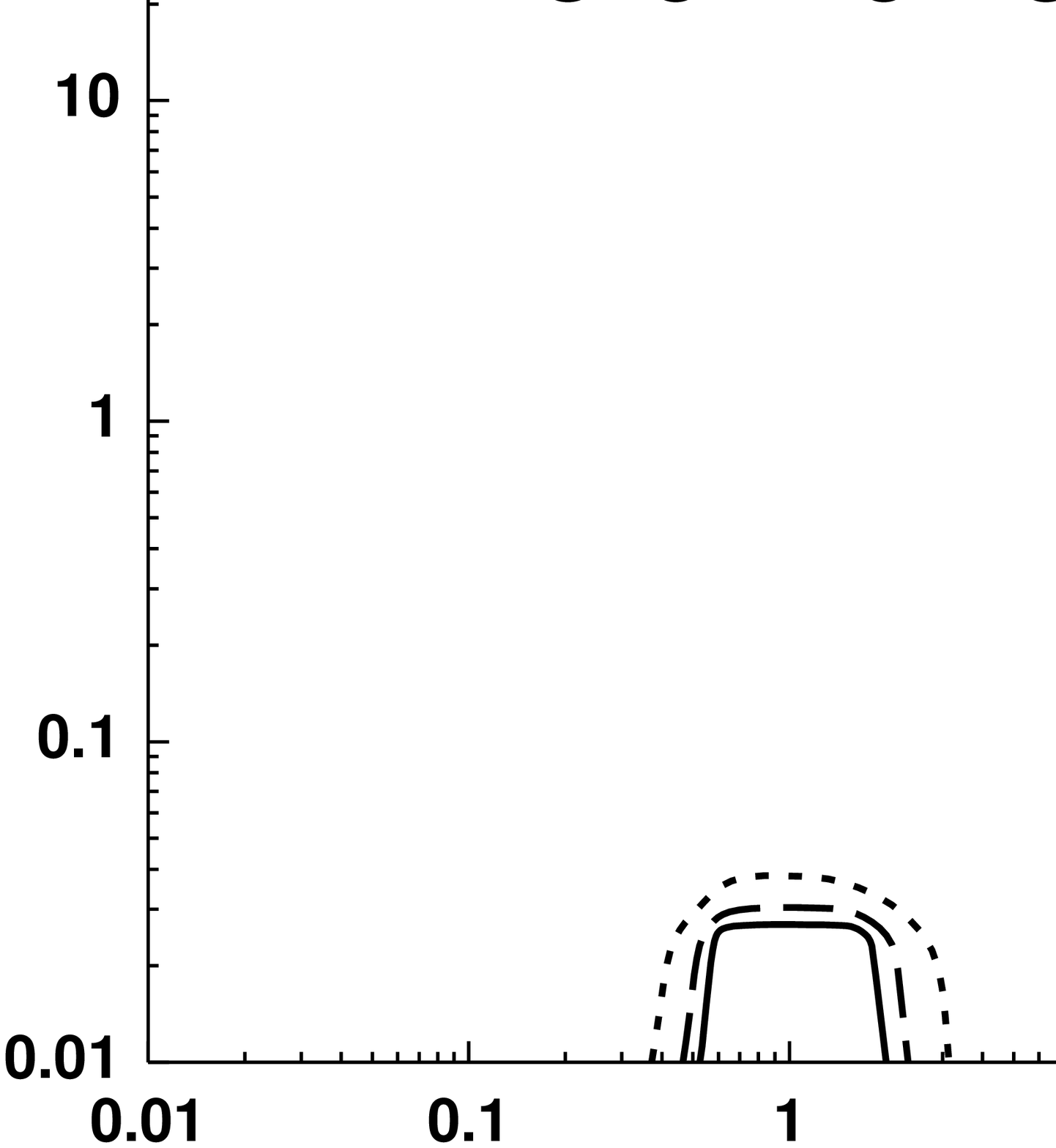,width=3.5cm}
\vglue -3.1cm \hglue 9.6cm \epsfig{file=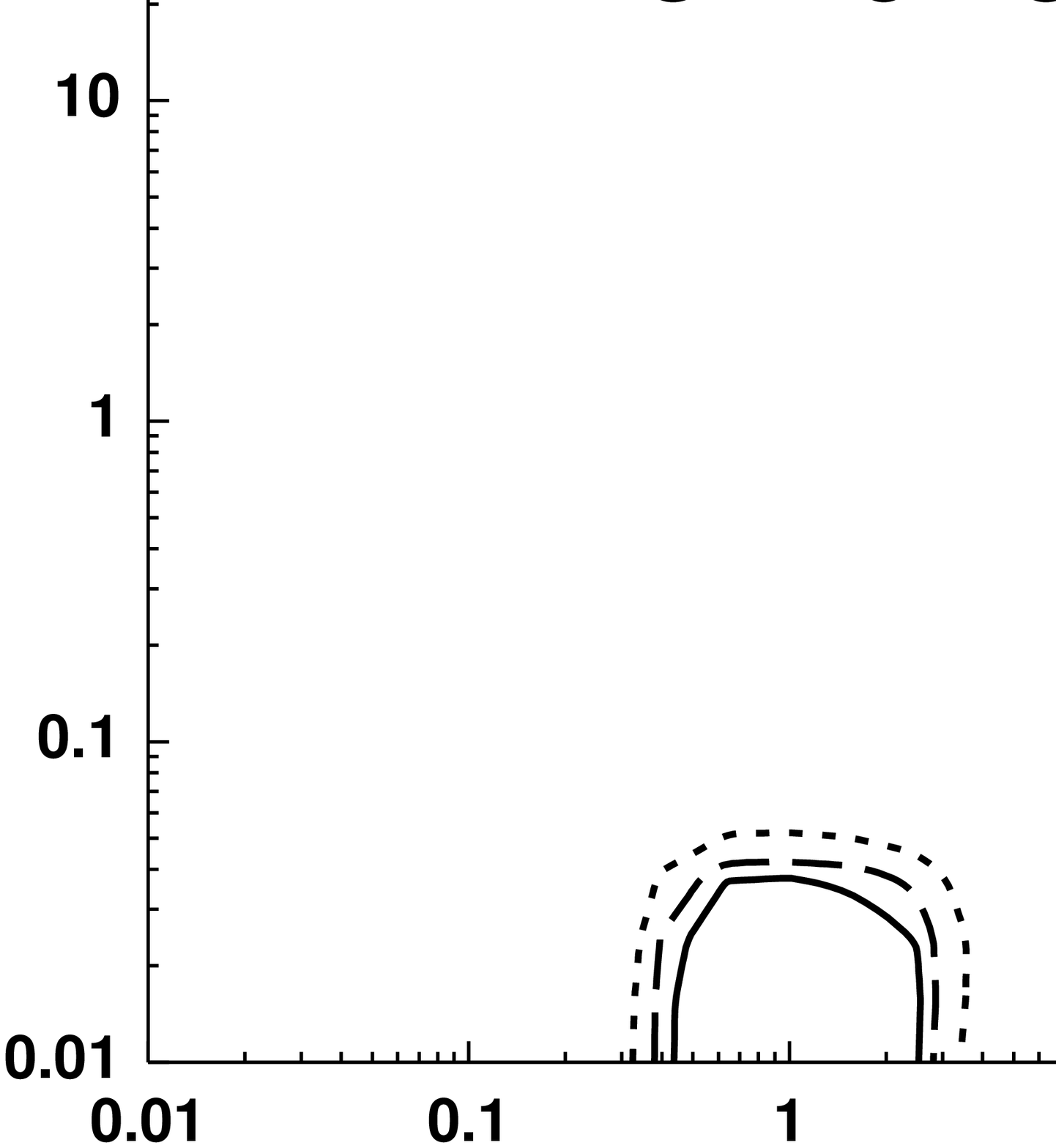,width=3.5cm}

\vglue 1.8cm
\hglue -1cm 
\epsfig{file=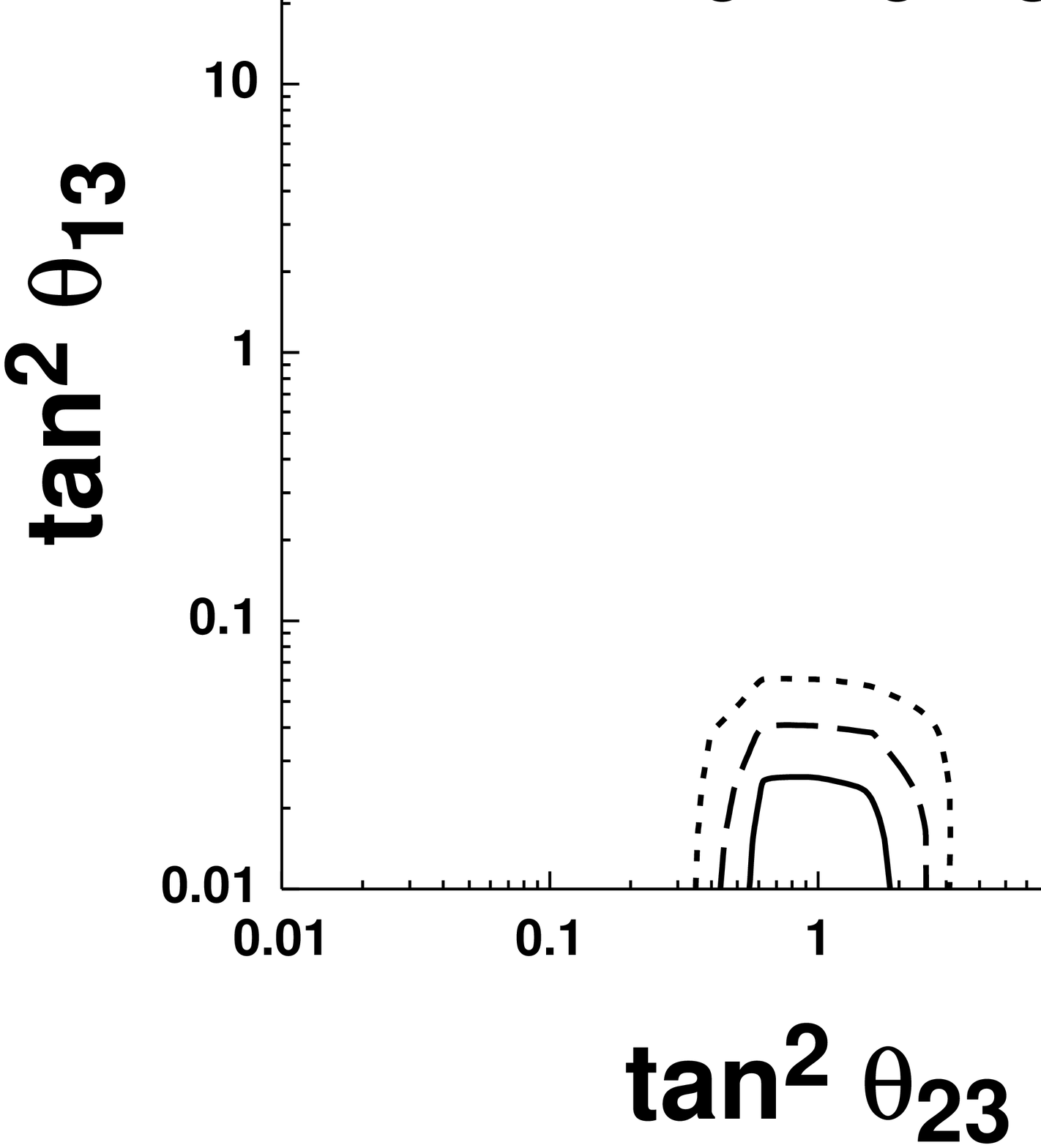,width=3.5cm}
\vglue -3.1cm \hglue 4.3cm \epsfig{file=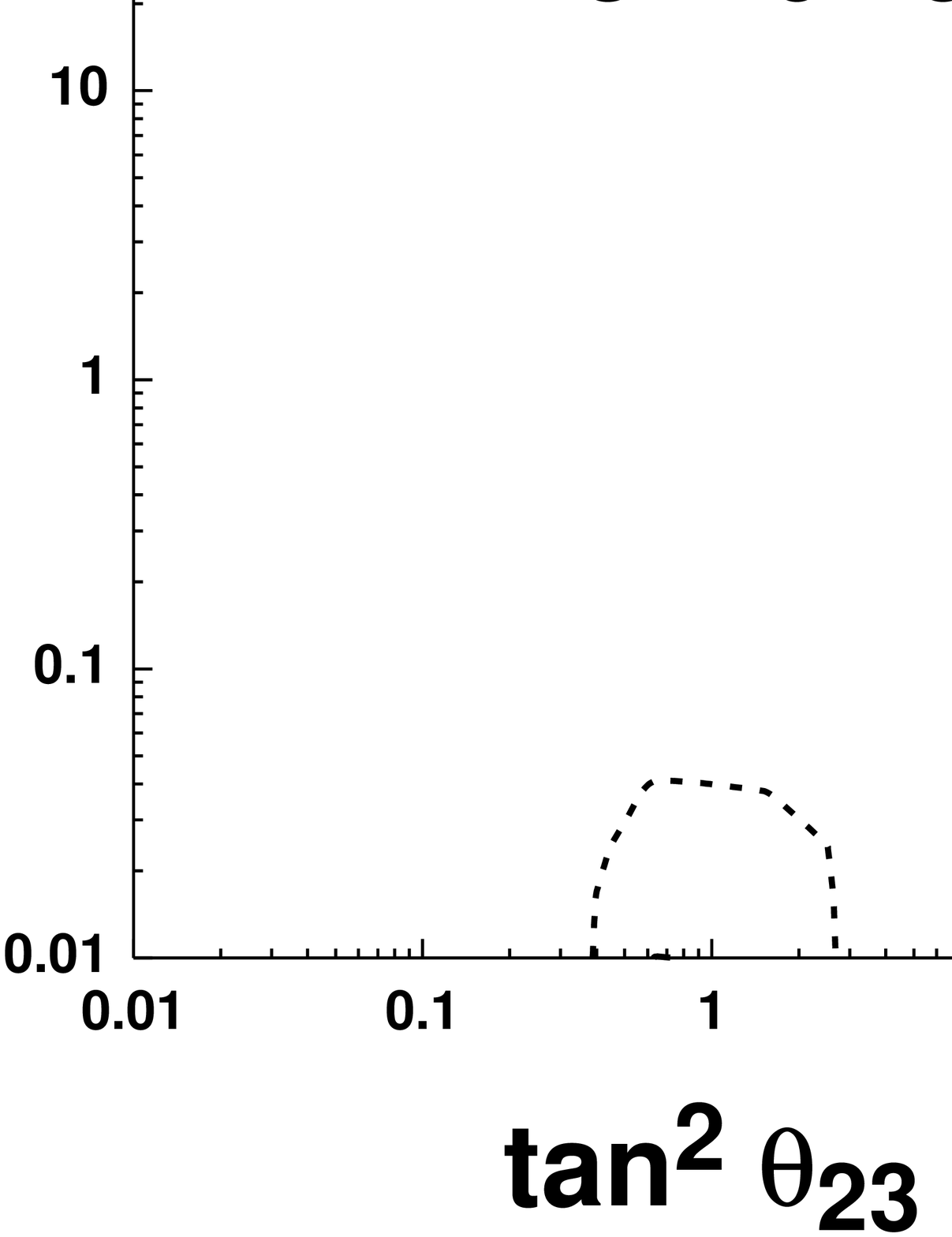,width=3.5cm}
\vglue -3.1cm \hglue 9.6cm \epsfig{file=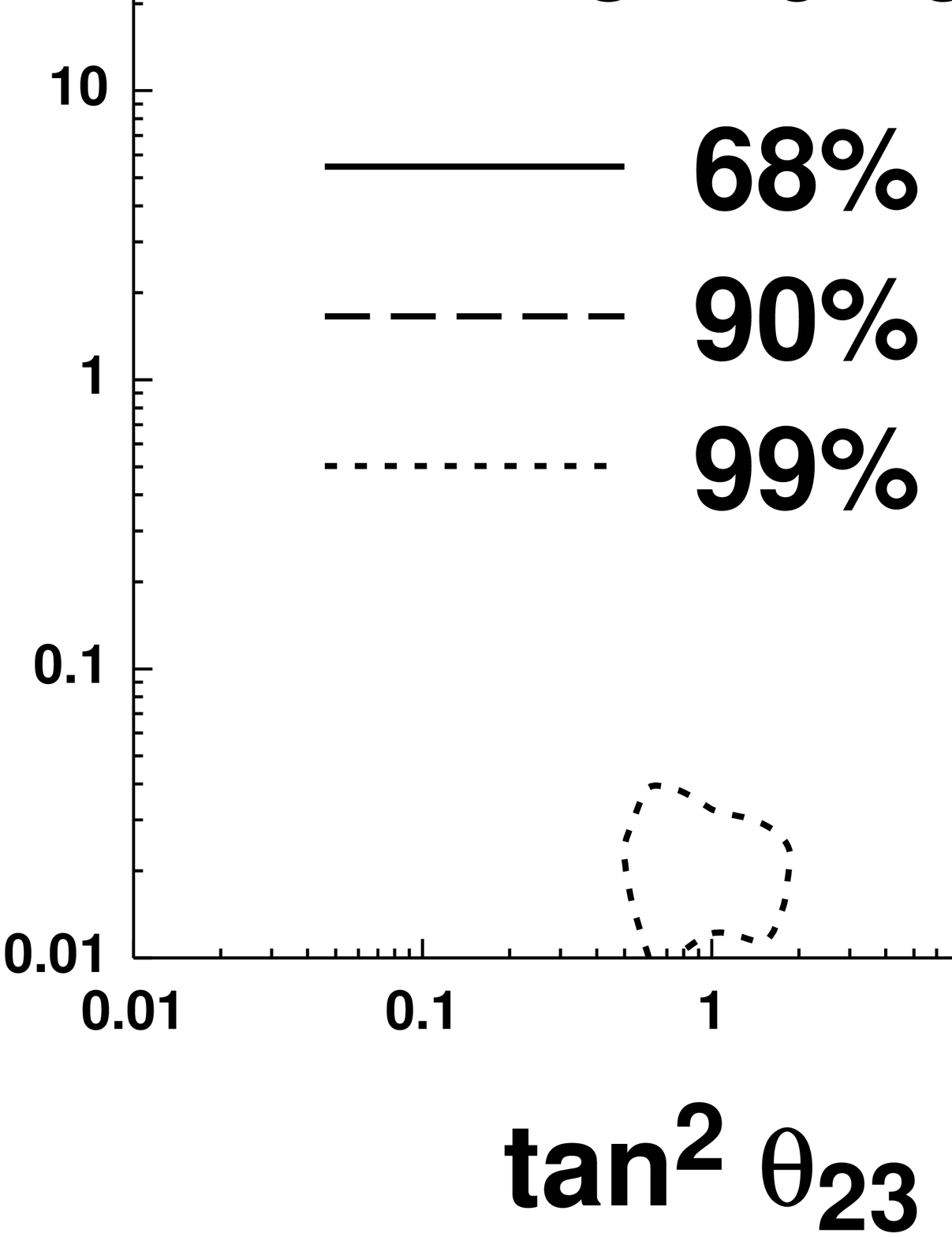,width=3.5cm}

\hglue 1cm\parbox{13cm}
{Fig. 3~~The solid, dashed,
dotted lines represent the allowed region at
68 \% CL, 90 \% CL, 99 \% CL, respectively for
scenario (a) in Fig. 1.}
\newpage
\vglue 2.5cm
\hglue -1cm 
\epsfig{file=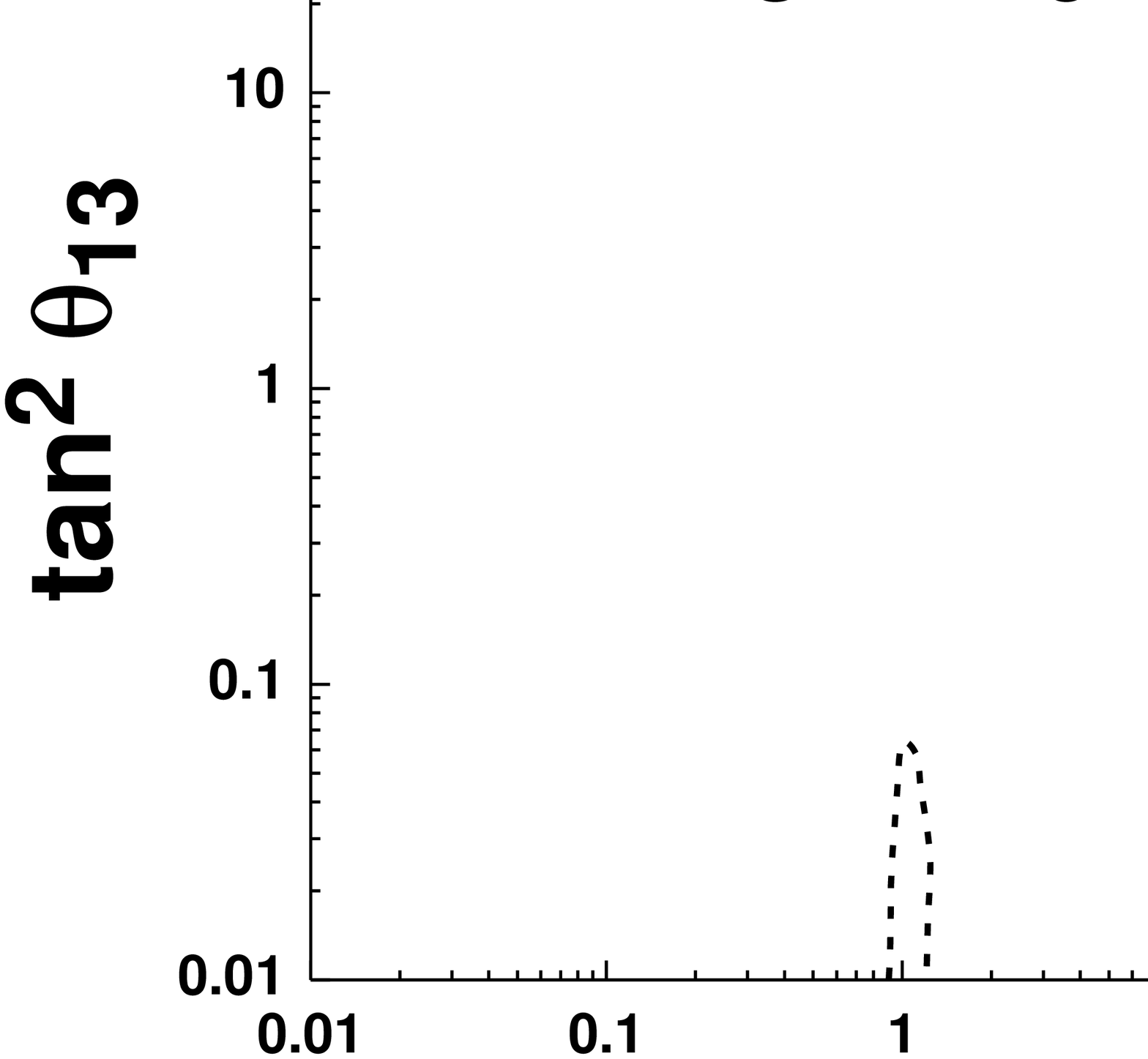,width=3.5cm}
\vglue -3.1cm \hglue 4.3cm \epsfig{file=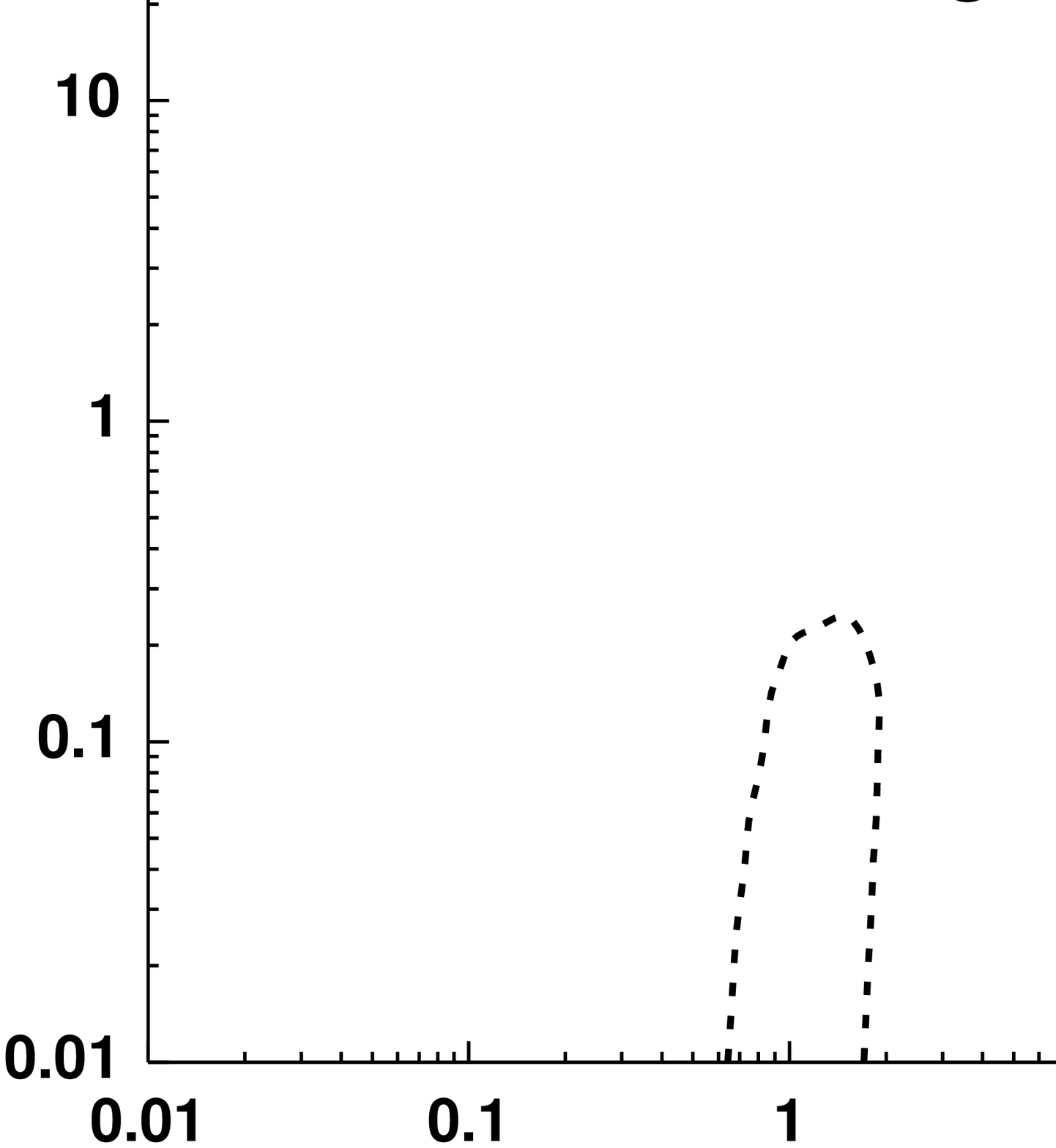,width=3.5cm}
\vglue -3.1cm \hglue 9.6cm \epsfig{file=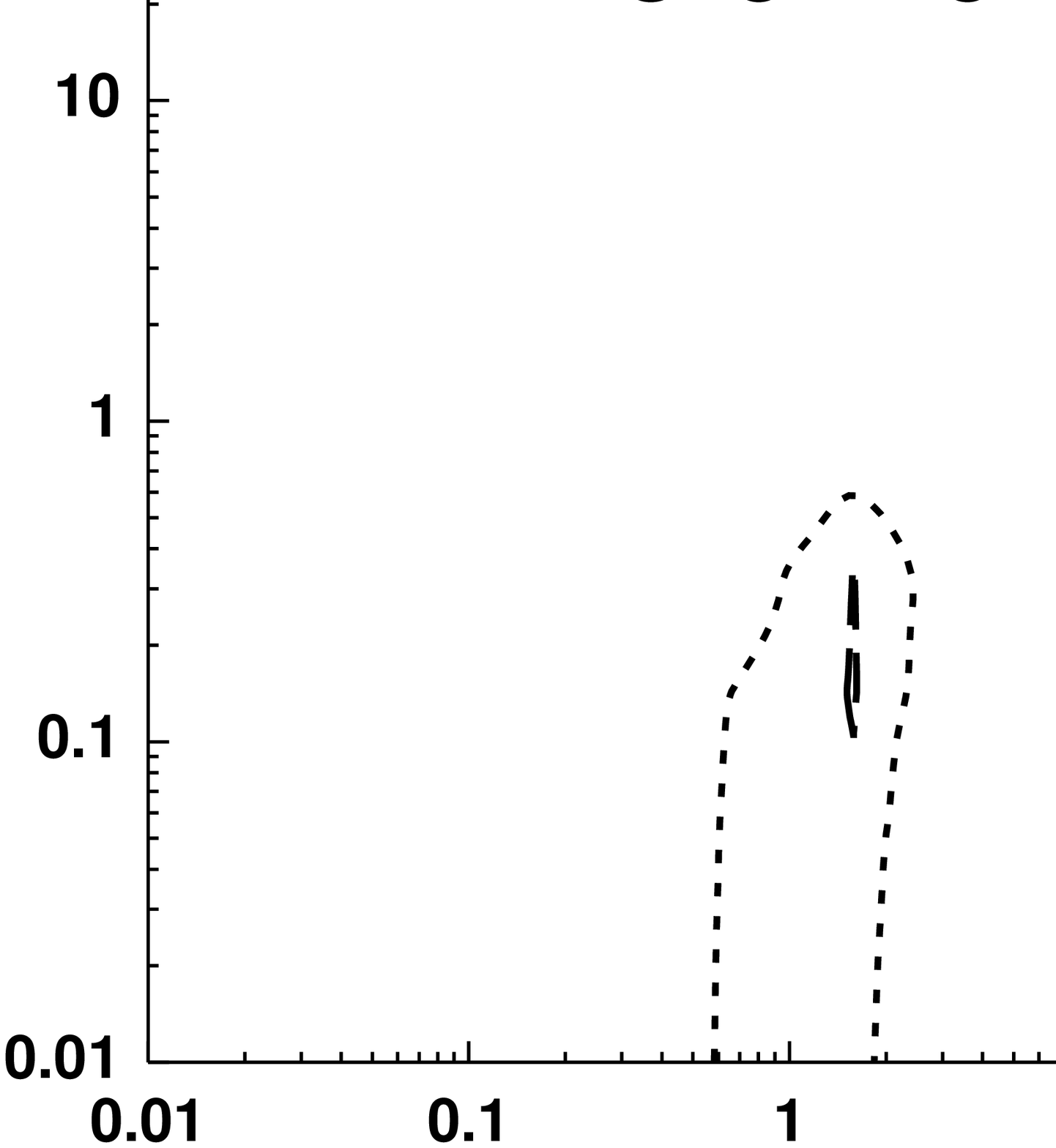,width=3.5cm}

\vglue 1.8cm
\hglue -1cm 
\epsfig{file=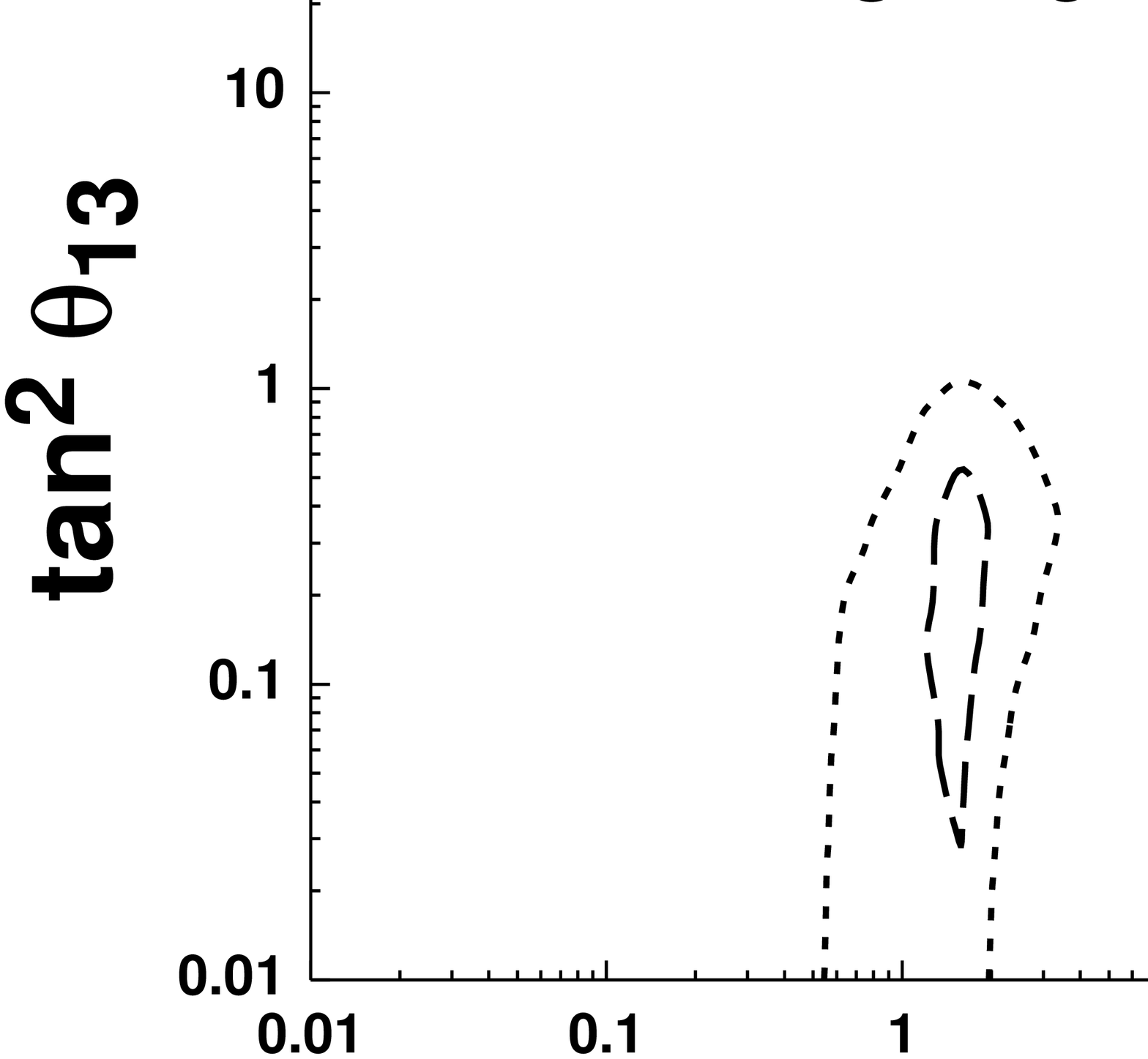,width=3.5cm}
\vglue -3.1cm \hglue 4.3cm \epsfig{file=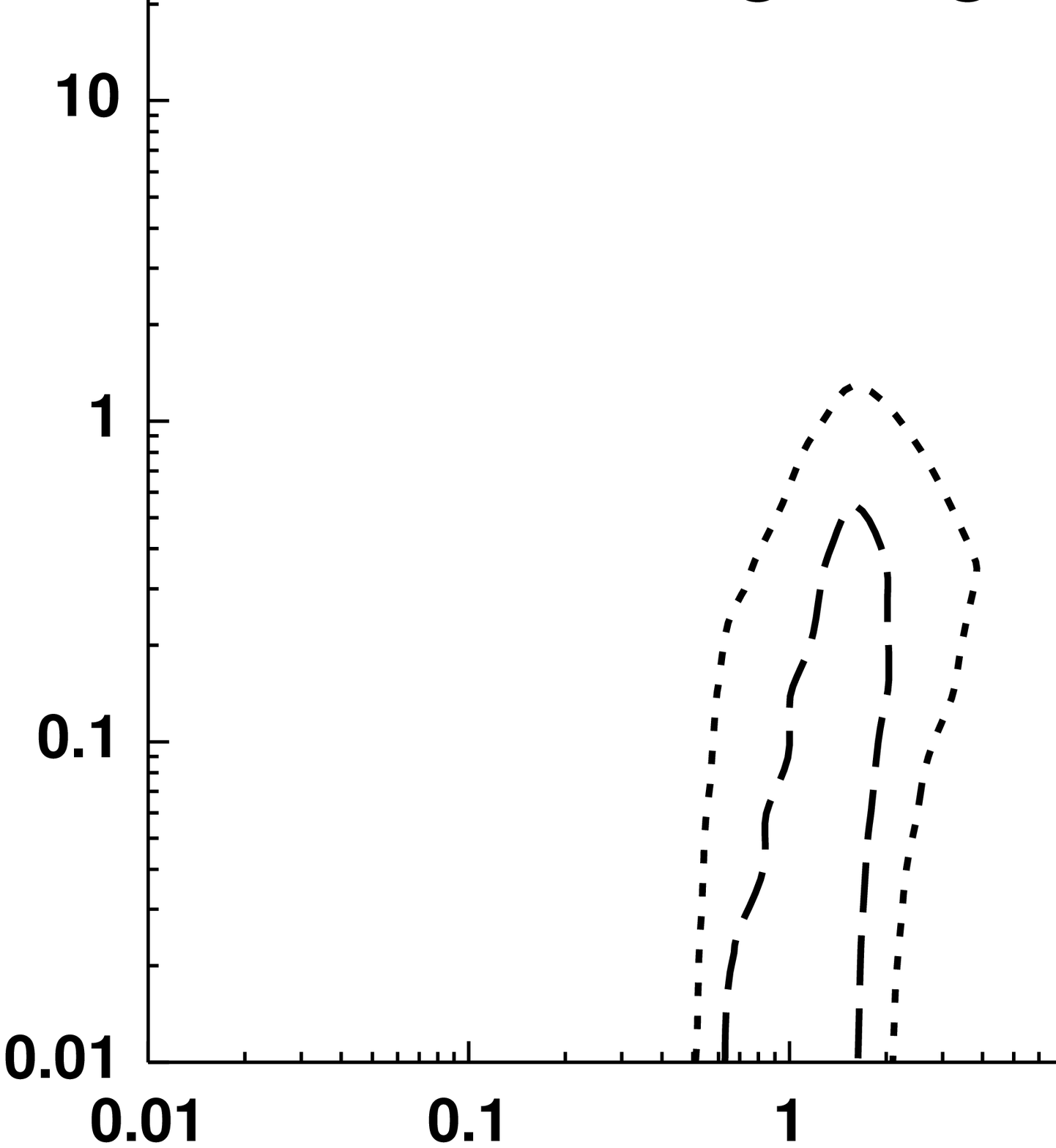,width=3.5cm}
\vglue -3.1cm \hglue 9.6cm \epsfig{file=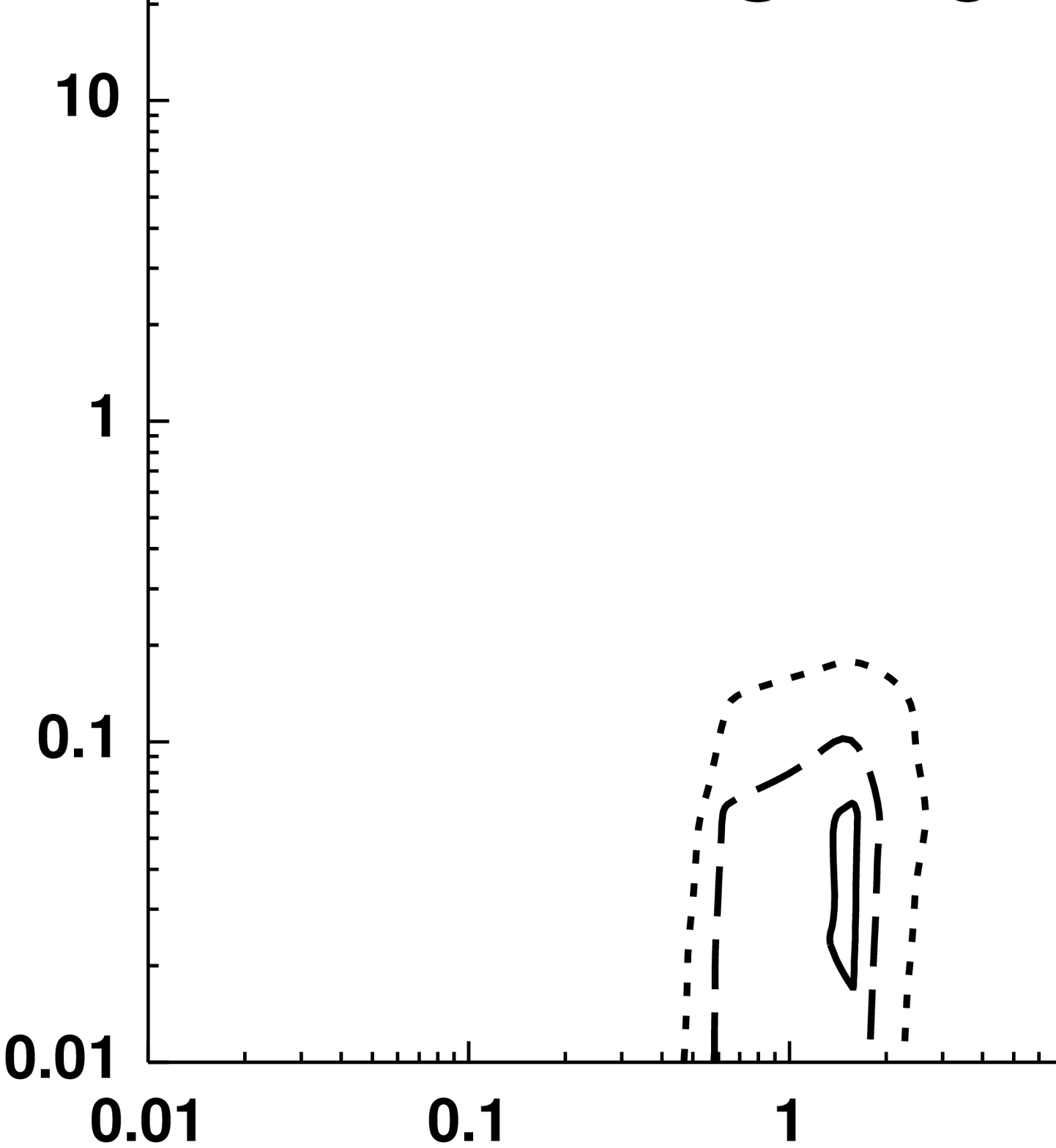,width=3.5cm}

\vglue 1.8cm
\hglue -1cm 
\epsfig{file=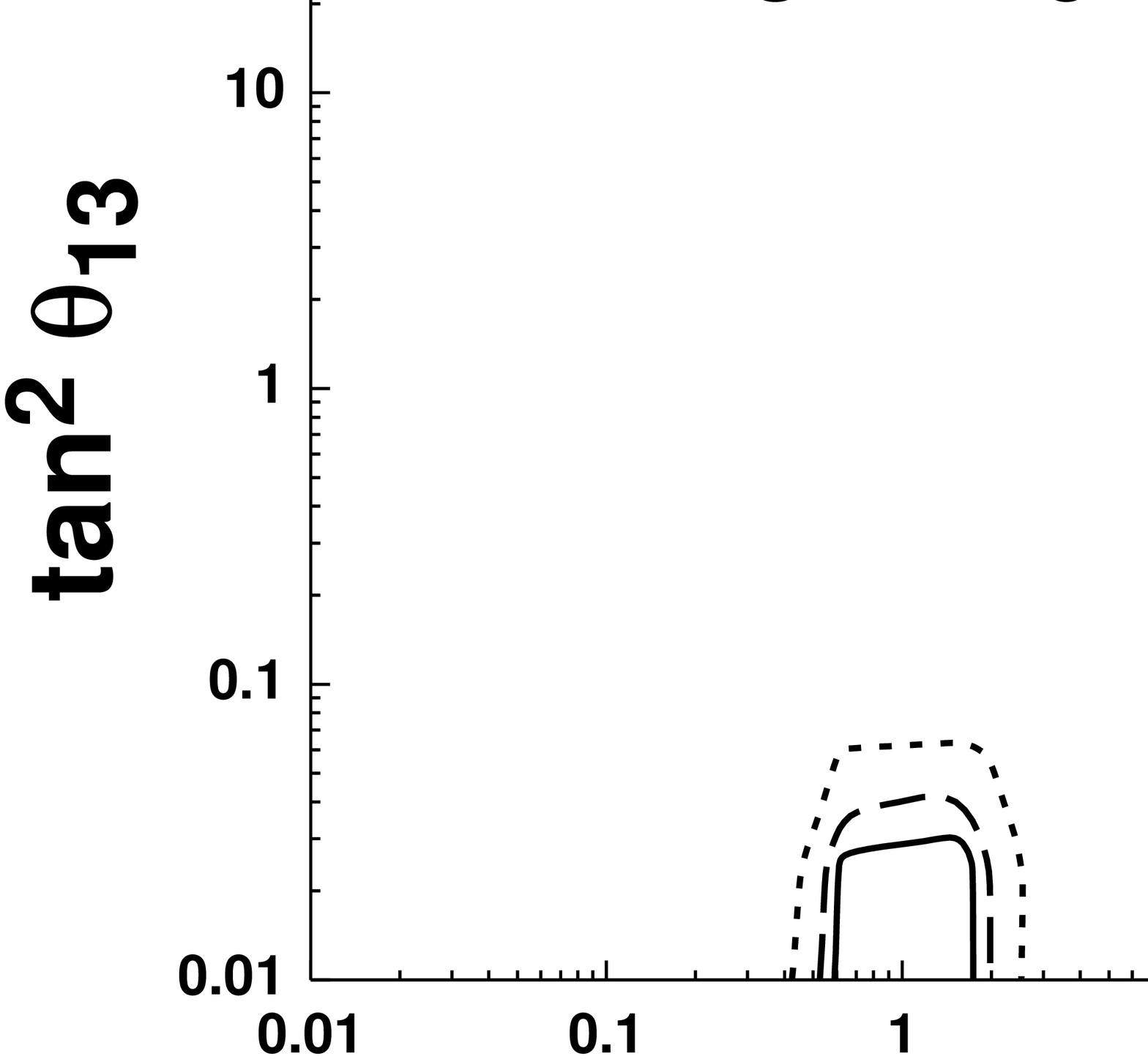,width=3.5cm}
\vglue -3.1cm \hglue 4.3cm \epsfig{file=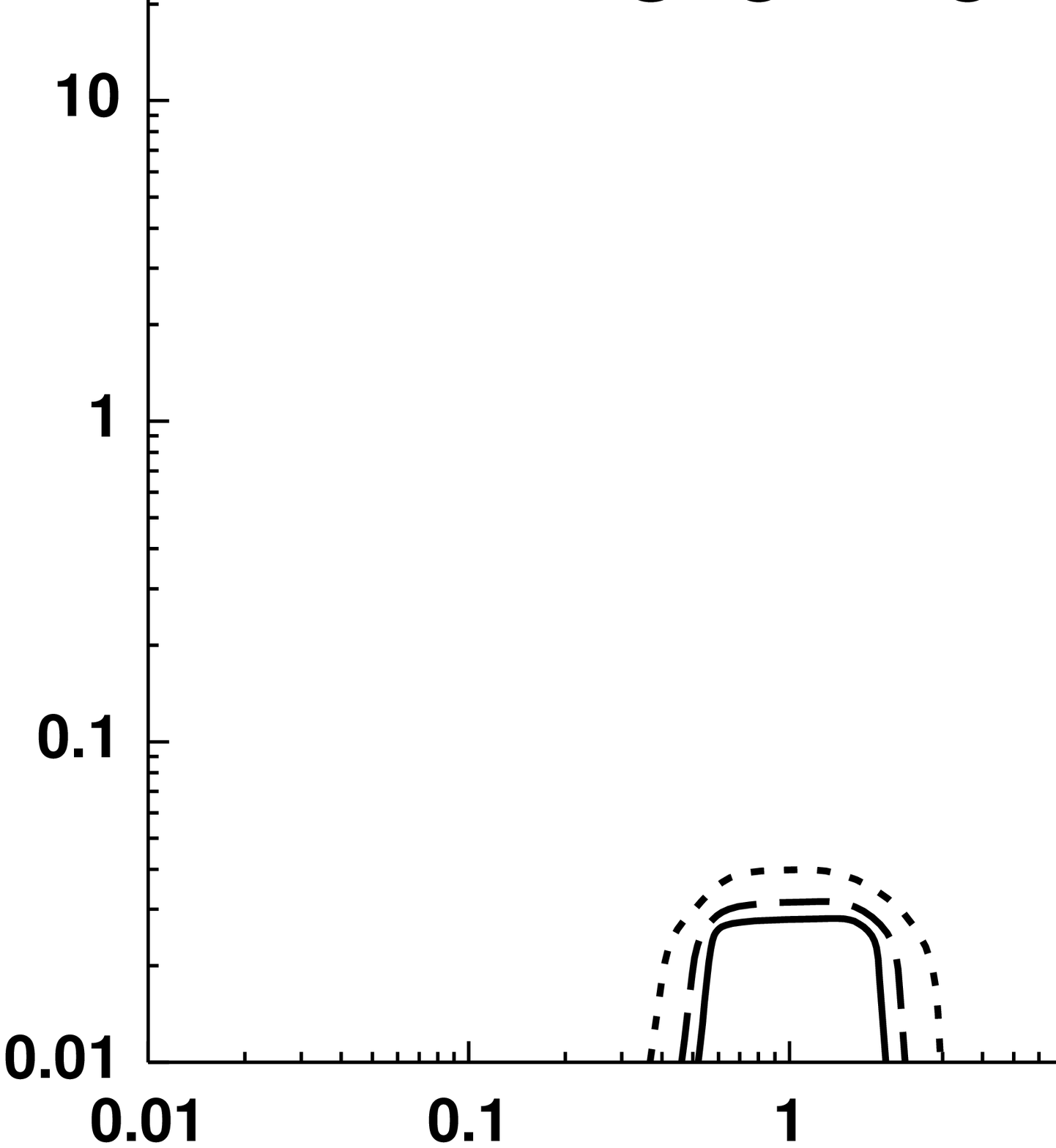,width=3.5cm}
\vglue -3.1cm \hglue 9.6cm \epsfig{file=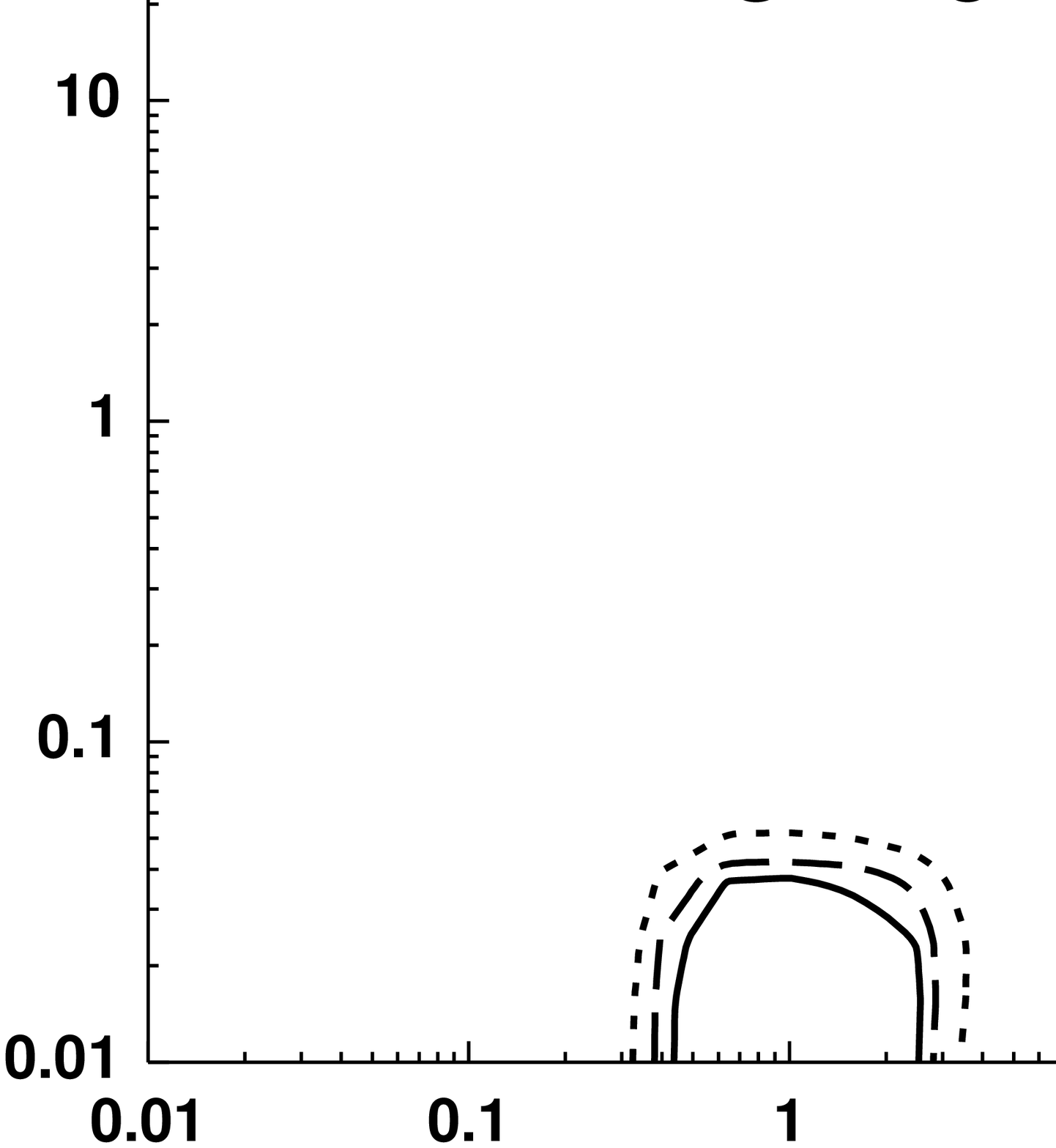,width=3.5cm}

\vglue 1.8cm
\hglue -1cm 
\epsfig{file=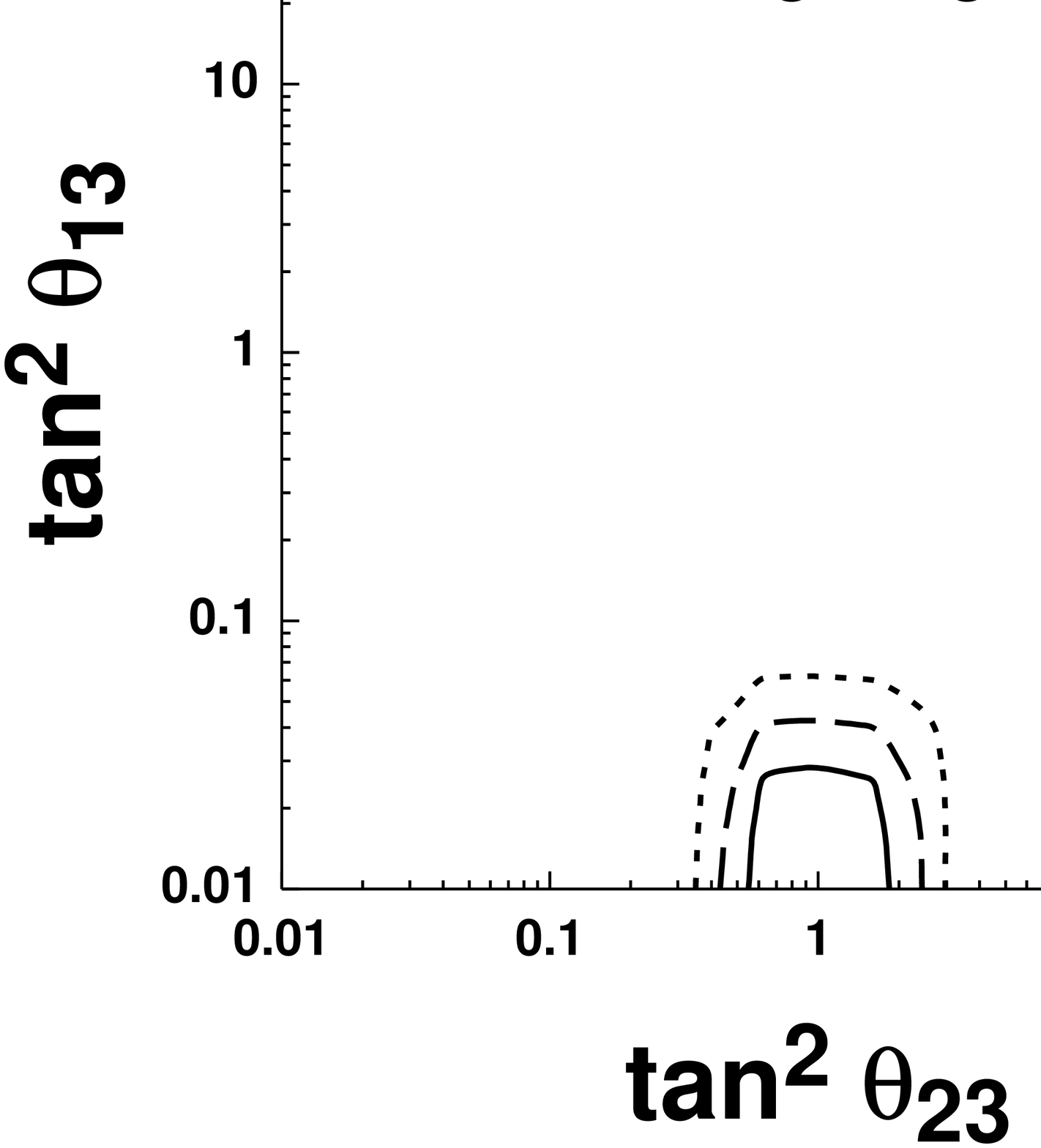,width=3.5cm}
\vglue -3.1cm \hglue 4.3cm \epsfig{file=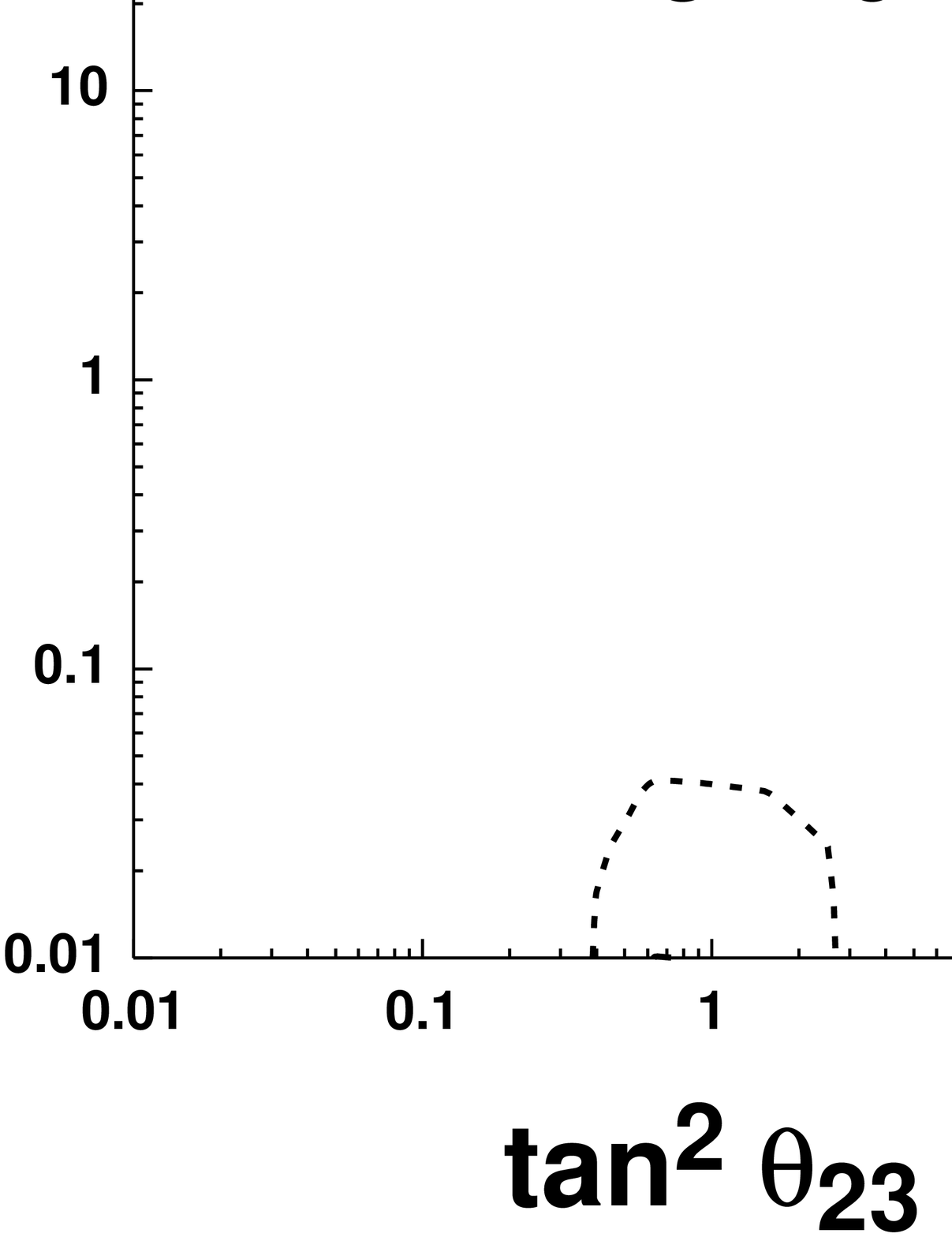,width=3.5cm}
\vglue -3.1cm \hglue 9.6cm \epsfig{file=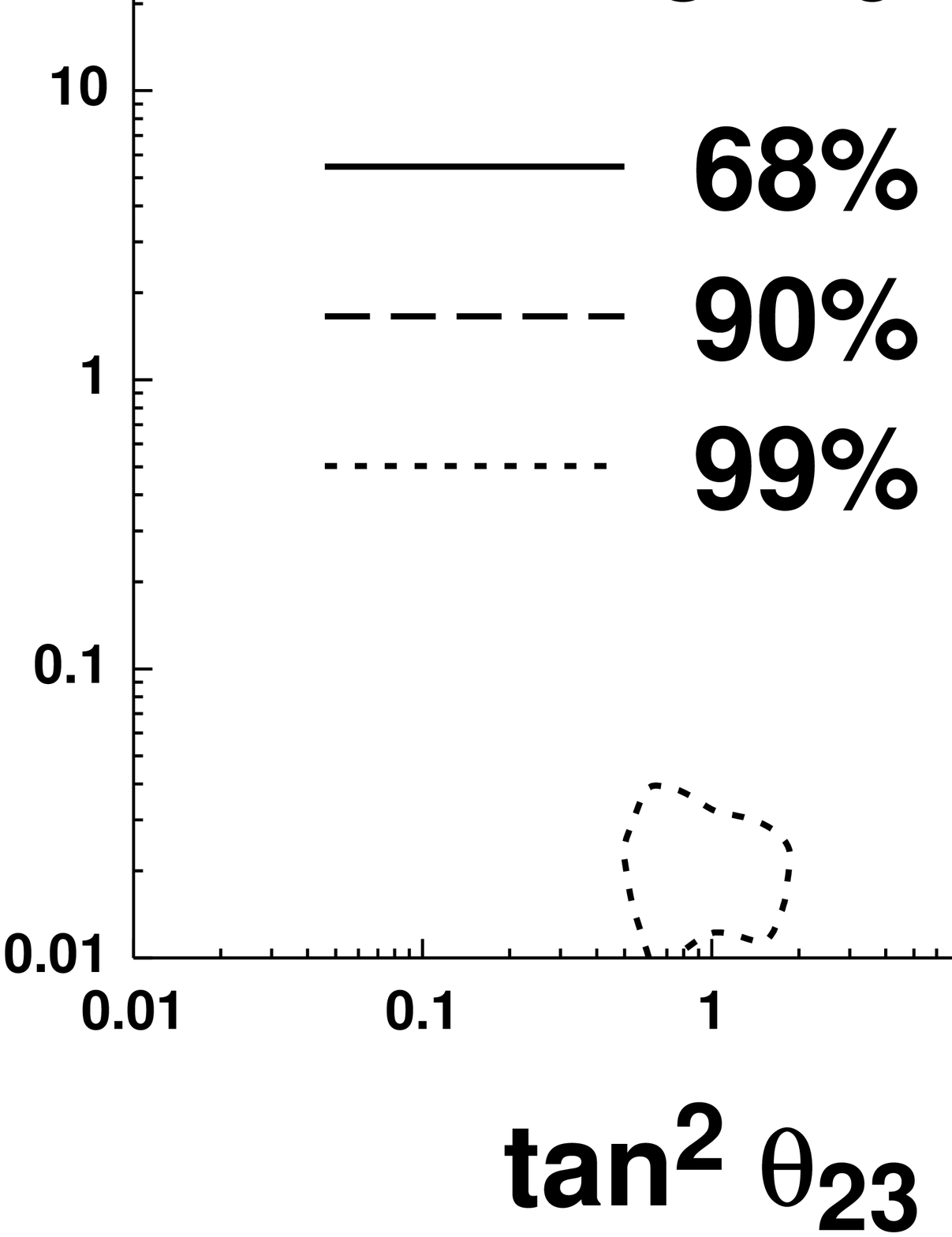,width=3.5cm}

\hglue 1cm\parbox{13cm}
{Fig. 4~~As in Fig. 3, but the scenario (b) in Fig. 1 is
assumed.}
\section{The implications to the long baseline experiments}
In the present case with mass scale hierarchy the oscillation
probability $P(\nu_\mu\rightarrow\nu_e)$ in vacuum is given by
\begin{eqnarray}
P(\nu_\mu\rightarrow\nu_e)=4|U_{e3}|^2|U_{\mu 3}|^2
\sin^2\left({\Delta m^2_{31}L \over 4E}\right)
=s_{23}^2\sin^22\theta_{13}\sin^2\left({m^2L \over 4E}\right),
\nonumber
\end{eqnarray}
so that we observe that the factor $4|U_{e3}|^2|U_{\mu 3}|^2$
corresponds to $\sin^22\theta$ in the two flavor framework.  For each
$m^2$ in Figs. 3 and 4 the maximal allowed values of
$4|U_{e3}|^2|U_{\mu 3}|^2$ have been evaluated and are shown in
Figs. 5(a) and 5(b) in the two flavor plot ($\sin^22\theta$, $\Delta
m^2$), respectively, where the regions which can be probed by
K2K$^{23}$, MINOS$^{21}$ and KamLAND$^{26,25}$ are also given.  The
only promising channel for the K2K experiment is
$\nu_\mu\leftrightarrow\nu_\mu$ disappearance, while MINOS might have
a chance to see $\nu_\mu\rightarrow\nu_e$ appearance and KamLAND may
be also able to observe $\bar\nu_e\leftrightarrow\bar\nu_e$
disappearance.

\hglue -1.5cm\epsfig{file=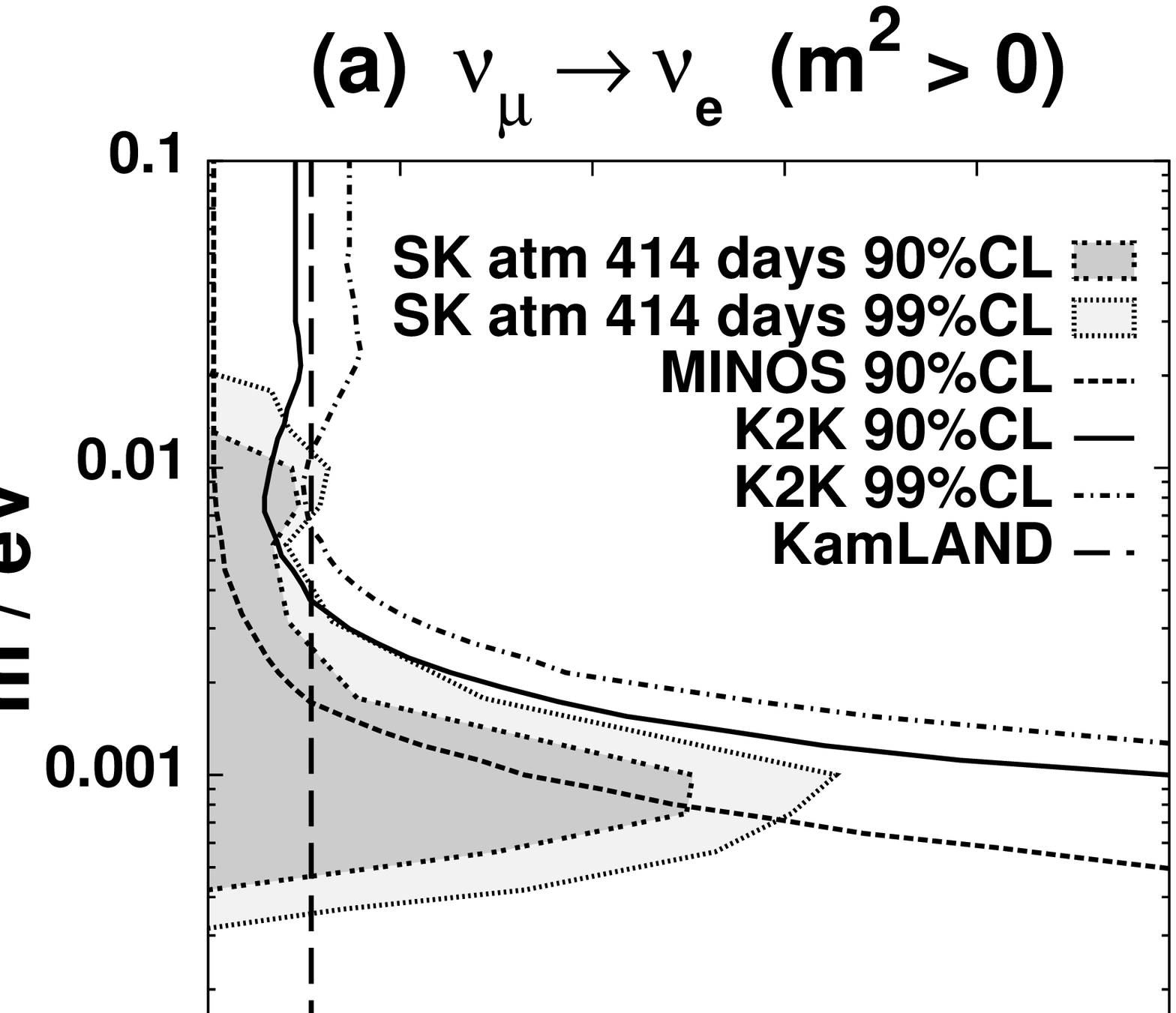,width=8cm}
\vglue -7cm\hglue 7.5cm
\epsfig{file=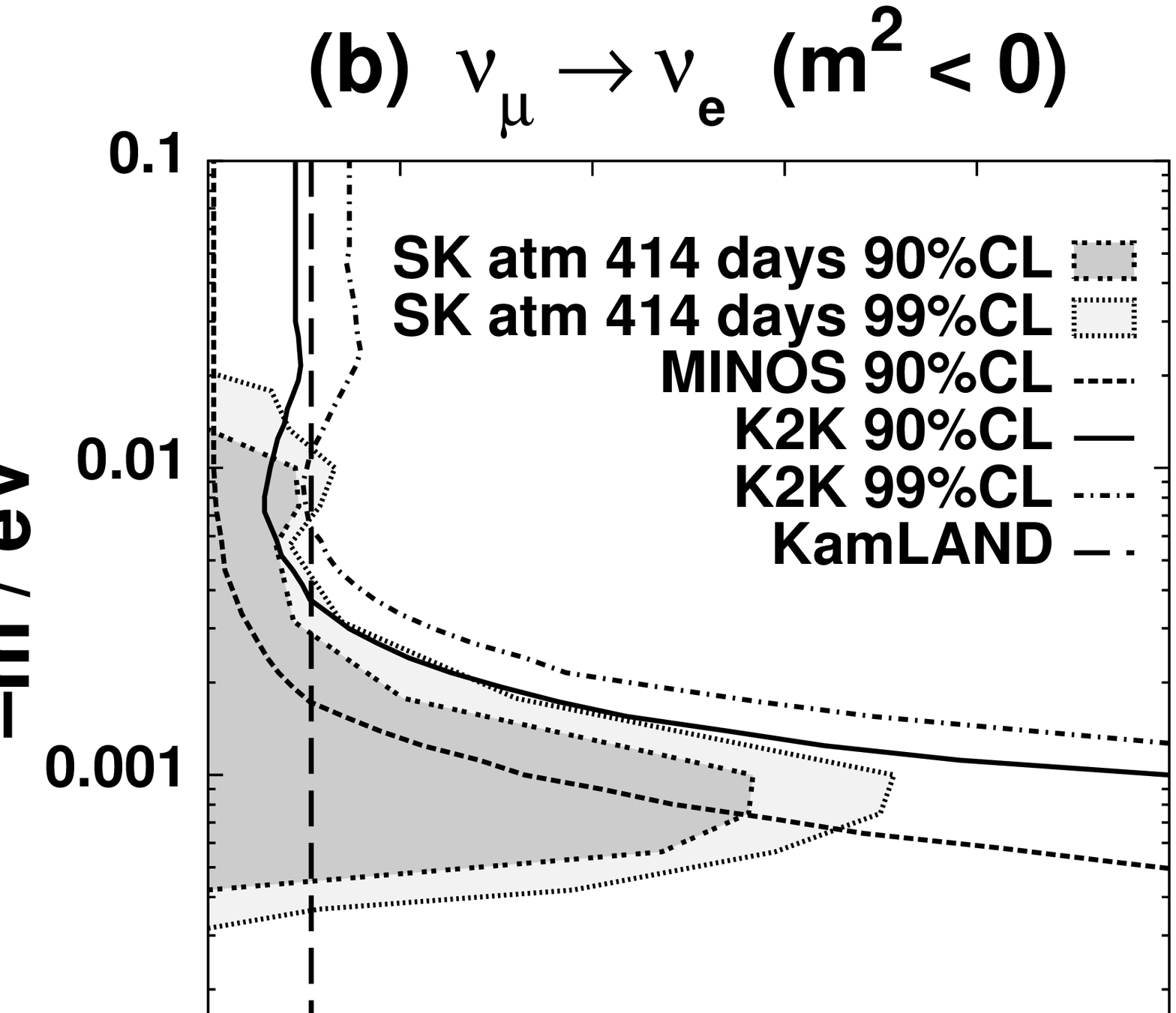,width=8cm}

\vglue 3cm\hglue 0.5cm\parbox{13cm} {Fig. 5\\ Shadowed
regions are allowed from the Superkamiokande atmospheric neutrino data
for the scenario (a) (Fig. 5(a)) or (b) (Fig. 5(b)) in Fig. 1,
respectively.  The region which can be probed by the long baseline
experiments (K2K, MINOS, KamLAND) is the right side of each line.}
\vglue 4cm
\section{The threefold maximal mixing model$^{22,14,13}$}
Using (\ref{eqn:sch}) and (\ref{eqn:sch2}), and redefining
$\nu_e\rightarrow e^{i\delta}\nu_e$ we have
\begin{eqnarray}
&{ }&i {d \over dx} \left( \begin{array}{c} \nu_e (x) \\ \nu_{\mu}(x) \\ 
\nu_{\tau}(x)
\end{array} \right)\nonumber\\
&=&e^{i\theta_{23}\lambda_7}
\left[ e^{i\theta_{13}\lambda_5}
 {\rm diag} \left(0,0,\Delta E_{31} \right)
e^{-i\theta_{13}\lambda_5}
+{\rm diag} \left(A,0,0 \right) \right]
e^{-i\theta_{23}\lambda_7}
\left( \begin{array}{c} \nu_e (x) \nonumber\\
\nu_{\mu}(x) \\ \nu_{\tau}(x)
\end{array} \right)\\
&=&e^{i\theta_{23}\lambda_7}e^{i\theta_M\lambda_5}
\left[ {\Delta E_{31}+A \over 2}
 {\rm diag} \left(1,0,1\right)- {B \over 2}
 {\rm diag} \left(1,0,-1 \right)\right]
e^{-i\theta_M\lambda_5}
e^{-i\theta_{23}\lambda_7}\left( \begin{array}{c} \nu_e (x) \\
\nu_{\mu}(x) \\ \nu_{\tau}(x)
\end{array} \right),\nonumber\\
\label{eqn:sch3}
\end{eqnarray}
where $\theta_M$ is the mixing angle in matter given by
\begin{eqnarray}
\tan2\theta_M\equiv {\Delta E_{31}\sin2\theta_{13}
\over \Delta E_{31}\cos2\theta_{13}-A}\nonumber
\end{eqnarray}
as in the two flavor case$^{20,28}$ , and
\begin{eqnarray}
B\equiv \sqrt{\left(\Delta E_{31}\cos2\theta_{13}-A\right)^2
+\left(\Delta E_{31}\sin2\theta_{13}\right)^2}.\nonumber
\end{eqnarray}
If we assume that the density of the Earth is approximately
constant, then (\ref{eqn:sch3}) can be integrated as
\begin{eqnarray}
\left( \begin{array}{c} \nu_e (L) \\ \nu_{\mu}(L) \\ 
\nu_{\tau}(L)
\end{array} \right) = 
e^{i\theta_{23}\lambda_7}e^{i\theta_M\lambda_5}
 {\rm diag} \left(e^{-i\Phi},1,e^{-i\Psi}\right)
e^{-i\theta_M\lambda_5}
e^{-i\theta_{23}\lambda_7}
\left( \begin{array}{c} \nu_e (0) \\
\nu_{\mu}(0) \\ \nu_{\tau}(0)
\end{array} \right),\nonumber
\end{eqnarray}
where
\begin{eqnarray}
\left\{\begin{array}{c} \Phi \\
\Psi\end{array}\right\}
={L \over 2}\left(\Delta E_{31}+A\mp B\right).\nonumber
\end{eqnarray}
The oscillation probability is given by
\begin{eqnarray}
P(\nu_e\rightarrow\nu_e)&=&1-\sin^22\theta_M\sin^2
{\Phi-\Psi \over 2}\nonumber\\
P(\nu_e\rightarrow\nu_\mu)&=&s_{23}^2
\sin^22\theta_M\sin^2{\Phi-\Psi \over 2}\nonumber\\
P(\nu_\mu\rightarrow\nu_\mu)&=&1-s_{23}^4\sin^22\theta_M\sin^2
{\Phi-\Psi \over 2}-\sin^22\theta_{23}\left(
s_M^2\sin^2{\Phi \over 2}+c_M^2\sin^2{\Psi \over 2}\right)\nonumber\\
\label{eqn:prob}
\end{eqnarray}
One of the interesting applications is the behaviors of the
oscillation probability (\ref{eqn:prob}) in the three flavor maximal
mixing model$^{22,14,13}$, which is obtained by putting
$\theta_{12}=\theta_{23}=\pi/4$,
$\theta_{13}=\sin^{-1}\left(1/\sqrt{3}\right)$, $\delta=\pi/2$ in the
present parametrization.  If $|\Delta E_{31}|\ll A$ then we see from
(\ref{eqn:prob}) that the mixing angle $\theta_M$ becomes $\pi/2$, so
we get from (\ref{eqn:prob})
\begin{eqnarray}
P(\nu_e\rightarrow\nu_e)&\simeq&1\nonumber\\
P(\nu_\mu\rightarrow\nu_\mu)&\simeq&1-\sin^22\theta_{23}
\sin^2{\Phi \over 2}
=1-\sin^2{\Phi \over 2},
\label{eqn:prob2}
\end{eqnarray}
where we have used the fact
\begin{eqnarray}
\Phi\simeq{\Delta E_{31}L \over 2}\left(1+\cos2\theta_{13}\right) =
c_{13}^2\Delta E_{31}L={2 \over 3}\Delta E_{31}L.\nonumber
\end{eqnarray}
(\ref{eqn:prob2}) indicates that the mixing becomes pure
$\nu_\mu\leftrightarrow\nu_\tau$ with maximal mixing and
this is why a fit of this model to the data becomes reasonably good
for lower $\Delta m_{31}^2$$^9$.  On the other hand,
if $|\Delta E_{31}|\gg A$, then $\theta_M$ becomes $\theta_{13}$
and we find again from (\ref{eqn:prob})
\begin{eqnarray}
P(\nu_e\rightarrow\nu_e)&\simeq&1-{1 \over 2}\sin^22\theta_{13}=
{5 \over 9}\nonumber\\
P(\nu_\mu\rightarrow\nu_\mu)&\simeq&1-{1 \over 2}s_{23}^4\sin^22\theta_{13}
-\sin^22\theta_{23}\left(s_{13}^2\sin^2{AL \over 3}
+{1 \over 2}c_{13}^2\right)\nonumber\\
&=&{5 \over 9}-{1 \over 3}\sin^2{AL \over 3},\nonumber
\end{eqnarray}
where we have used
\begin{eqnarray}
\Phi\simeq{\Delta AL \over 2}\left(1+\cos2\theta_{13}\right) =
c_{13}^2AL={2 \over 3}AL.\nonumber
\end{eqnarray}
Possibility with larger $\Delta m_{31}^2$ is excluded by the CHOOZ
result$^2$, but a fit of this model to the atmospheric neutrino data
themselves is reasonably good because of the factor
$\sin^2\left(AL/3\right)$ accounts for the zenith angle dependence to
some extent$^{6,9}$.  These two behaviors can be confirmed numerically
as is shown in Fig. 6 where the value of $\chi^2_{\rm CHOOZ}$ is also
plotted.  Thus we see the analytic formulae for constant density
explain the qualitative behaviors of the oscillation probability well.
The threefold maximal mixing model fits to the data of atmospheric
neutrinos and the CHOOZ experiment for $m^2\simeq 8\times
10^{-4}$eV$^2$ $^9$.

\epsfig{file=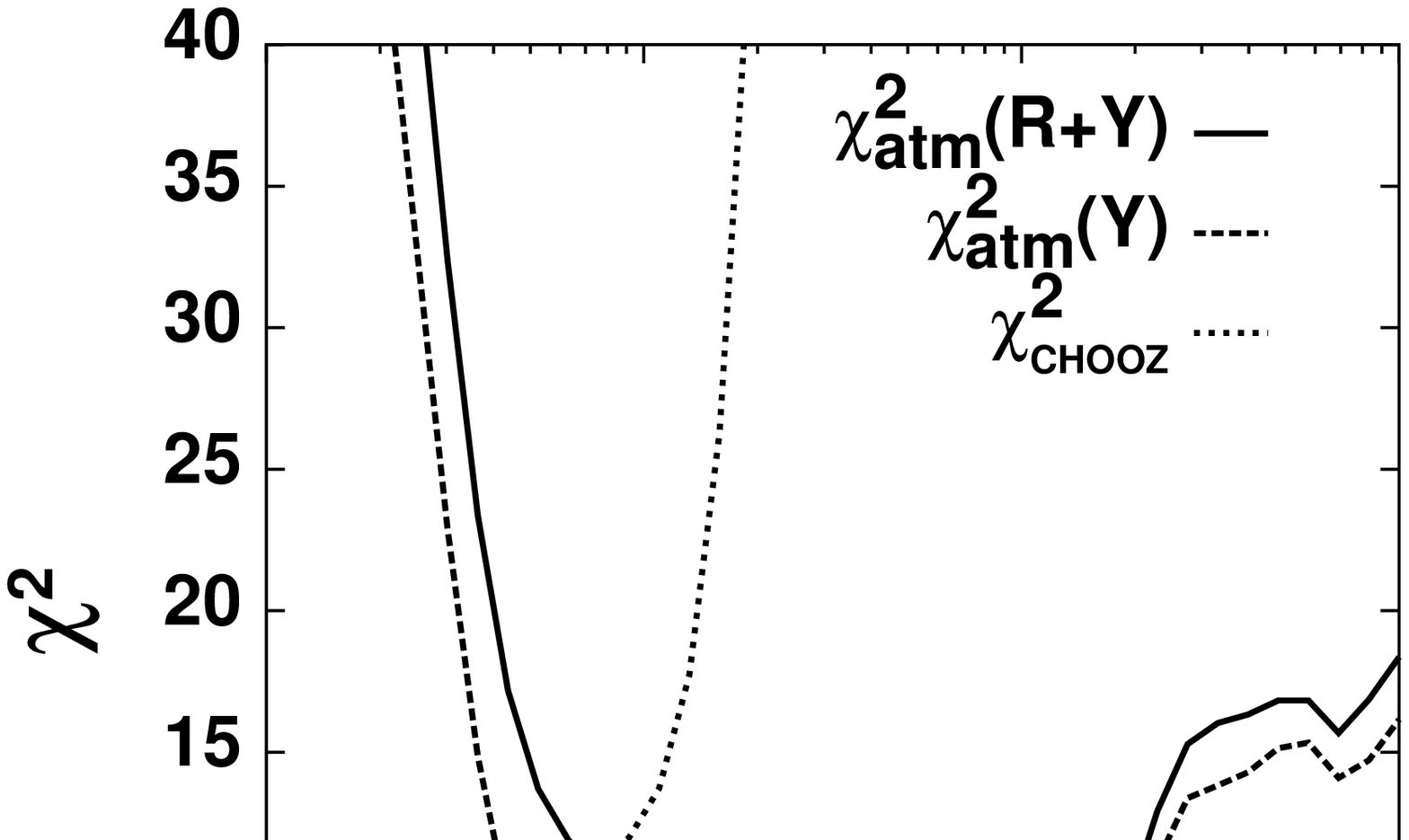,width=8cm}
\vglue -4.0cm\hglue 8cm
\parbox{5cm}{Fig. 6\\ \fussy{\small $\chi^2$ fit as a function of
$\Delta m^2$ to the Superkamiokande atmospheric data for the threefold
maximal mixing model.  The solid line includes both $R$ and up-down
asymmetries whereas the dashed line includes only the up-down
asymmetries (\# of degrees of freedom = 6--1=5 for R+Y and =4--1=3 for
Y only).  The $\chi^2$ for the CHOOZ reactor data (dotted line; \# of
degrees of freedom = 12--1=11) is also shown.}}

\section{Conclusions}
I have presented a result on the three flavor analysis of the
Superkamiokande atmospheric neutrino data for 414 days.  If the mass
squared difference is smaller than $1\times 10^{-3}$eV$^2$ then there
can be large mixing between $\nu_\mu$ and $\nu_e$.  It was shown that
K2K will be able to see only $\nu_\mu\leftrightarrow\nu_\mu$
disappearance, while there may be a chance for MINOS (KamLAND) to see
$\nu_\mu\rightarrow\nu_e$ appearance
($\bar\nu_e\leftrightarrow\bar\nu_e$ disappearance), respectively.
The threefold maximal mixing model gives a reasonable fit to the
atmospheric neutrino and CHOOZ data for $\Delta m^2\sim1\times
10^{-3}$eV$^2$.  It should be emphasized that the matter effects are
important in the analysis of atmospheric neutrinos. \vglue 0.5cm

\noindent
{\bf Acknowledgement}
\vglue 0.5cm

The author would like to thank H. Minakata and E. Lisi for
discussions, R. Foot and R.R. Volkas for collaboration and
discussions.  This research was supported in part by a Grant-in-Aid
for Scientific Research of the Ministry of Education, Science and
Culture, \#09045036, \#10140221, \#10640280.
\section{References}
\vspace{1pc}

\re
1. B. Ackar et al., 1995, Nucl. Phys. {\bf B434},  503.
\re
2. M. Apollonio et al., 1998, Phys. Lett. {\bf B420}, 397.
\re
3. See, e.g., J.N. Bahcall, R. Davis, Jr., P. Parker, A. Smirnov, R. Ulrich
eds., 1994, {\sl SOLAR NEUTRINOS: the first thirty years}
Reading, Mass., Addison-Wesley and references therein.
\re
4. V. Barger, K. Whisnant, S. Pakvasa and R.J.N. Phillips, 1980, Phys. Rev. {\bf D22}, 2718.
\re
5. G.L. Fogli, E. Lisi, D. Montanino and G. Scioscia, 1997, Phys. Rev. {\bf D55}, 4385.
\re
6. G. L. Fogli, E. Lisi, A. Marrone, and D. Montanino, 1998, Phys. Lett. {\bf B425}, 341.
\re
7. G. L. Fogli, E. Lisi, A. Marrone, and G. Scioscia, 1998, hep-ph/9808205.
\re
8. R. Foot, R. R. Volkas and O. Yasuda, 1998, Phys. Rev. {\bf D58}, 13006.
\re
9. R. Foot, R. R. Volkas and O. Yasuda, 1998, Phys. Lett. {\bf B433}, 82.
\re
10. Y. Fukuda et al., 1994, Phys. Lett. {\bf B335}, 237.
\re
11. Y. Fukuda et al., 1998, Phys. Lett. {\bf B433}, 9; 1998, hep-ex/9805006.
\re
12. Y. Fukuda et al., 1998, hep-ex/9807003.
\re
13. C. Giunti, C. W. Kim and J. D. Kim, 1995, Phys. Lett. {\bf B352}, 357.
\re
14. P. F. Harrison, D. H. Perkins and W. G. Scott, 1995, Phys. Lett. {\bf B349}, 137; 1997, Phys. Lett. {\bf B396}, 186.
\re
15. K. S. Hirata et al., 1992, Phys. Lett. {\bf B280}, 146.
\re
16. ICARUS experiment, http://www.aquila.infn.it/icarus/.
\re
17. T. Kajita, Talk at {\it XVIII International Conference on Neutrino
Physics and Astrophysics}, June, 1998, Takayama, Japan\\
(http://www-sk.icrr.u-tokyo.ac.jp/nu98/scan/063/).
\re
18. E. Lisi, Talk at {\it New Era in Neutrino Physics}, June, 1998,
Tokyo Metropolitan University, Japan\\
(http://musashi.phys.metro-u.ac.jp/neutrino/june12/e-lisi/).
\re
19. Z. Maki, M. Nakagawa and S. Sakata, 1962, Prog. Theor. Phys. {\bf 28}, 870.
\re
20. S. P. Mikheyev and A. Smirnov, 1986, Nuovo Cim. {\bf 9C}, 17.
\re
21. MINOS experiment, http://www.hep.anl.gov/NDK/HyperText/numi.html.
\re
22. R. N. Mohapatra and S. Nussinov, 1995, Phys. Lett. {\bf B346}, 75.
\re
23. K. Nishikawa, preprint INS-Rep.924 (1992); Talk at {\it XVIII
International Conference on Neutrino Physics and Astrophysics},
June, 1998, Takayama, Japan (http://www-sk.icrr.u-tokyo.ac.jp/nu98/scan/093/).
\re
24. Particle Data Group, 1998, Eur. Phys. J. {\bf C3}, 1.
\re
25. F. Suekane, Talk at {\it New Era in Neutrino Physics}, June, 1998,
Tokyo Metropolitan University, Japan\\
(http://musashi.phys.metro-u.ac.jp/neutrino/june12/f-suekane/).
\re
26. A. Suzuki, Talk at {\it XVIII International Conference on Neutrino
Physics and Astrophysics}, June, 1998, Takayama, Japan\\
(http://www-sk.icrr.u-tokyo.ac.jp/nu98/scan/083/).
\re
27. G. S. Vidyakin et al., 1994, JETP Lett. {\bf 59}, 390.
\re
28. L. Wolfenstein, 1978, Phys. Rev. {\bf D17}, 2369.
\end{document}